\documentclass[reprint,superscriptaddress,nofootinbib,amsmath,amssymb,aps,nofootinbib]{revtex4-1}
\usepackage[sort&compress]{natbib}
\usepackage{graphicx}
\usepackage{dcolumn}
\usepackage{bm}
\usepackage{float}
\usepackage{ragged2e} 
\usepackage{changes}
\usepackage{color}
\usepackage{varioref}
\usepackage{hyperref}
\usepackage{cleveref}
\usepackage{amsmath}
\usepackage{stackengine}

\hypersetup{
    colorlinks,
    citecolor=black,
    filecolor=black,
    linkcolor=black,
    urlcolor=black
}
\usepackage{stackengine}
\stackMath
\usepackage{textcomp}
\begin{document}

\title{Thermodynamics of light management in near-field thermophotovoltaics}

\author{Georgia T. Papadakis}
\affiliation{ICFO-Institut de Ciencies Fotoniques, The Barcelona Institute of Science and Technology, 08860 Castelldefels (Barcelona), Spain}
\affiliation{Department of Electrical Engineering, Ginzton Laboratory, Stanford University, California, 94305, USA}
\author{Meir Orenstein}%
\affiliation{Department of Electrical Engineering, Technion-Israel Institute of Technology, 32000 Haifa, Israel}
\author{Eli Yablonovitch}
\affiliation{Department of Electrical Engineering \& Computer Sciences, University of California, Berkeley, California, 94720, USA}
\author{Shanhui Fan}
\email{shanhui@stanford.edu}
\affiliation{Department of Electrical Engineering, Ginzton Laboratory, Stanford University, California, 94305, USA}

\date{\today}

\begin{abstract}{We evaluate near-field thermophotovoltaic (TPV) energy conversion systems focusing in particular on their open-circuit voltage ($V_\mathrm{oc}$). Unlike previous analyses based largely on numerical simulations with fluctuational electrodynamics, here, we develop an analytic model that captures the physics of near-field TPV systems and can predict their performance metrics. Using our model, we identify two important opportunities of TPV systems operating in the near-field. First, we show analytically that enhancement of radiative recombination is a natural consequence of operating in the near-field. Second, we note that, owing to photon recycling and minimal radiation leakage in near-field operation, the PV cell used in near-field TPV systems can be much thinner compared to those used in solar PV systems. Since non-radiative recombination is a volumetric effect, use of a thinner cell reduces non-radiative losses per unit area. The combination of these two opportunities leads to increasingly large values of $V_\mathrm{oc}$ as the TPV vacuum gap decreases. Hence, although operation in the near-field was previously perceived to be beneficial for electrical power density enhancement, here, we emphasize that thin-film near-field TPVs are also significantly advantageous in terms of $V_\mathrm{oc}$ and consequently conversion efficiency \textit{as well as} power density. We provide numerical results for an InAs-based thin-film TPV that exhibits efficiency $>50\%$ at an emitter temperature as low as $1100$ K.}
\end{abstract}
\maketitle

\section{\label{sec:level1}Introduction}

\par{Global terawatt-scale energy needs call for renewable energy harvesting approaches operating close to thermodynamic limits \cite{Global_Energy}. Amongst light-based renewable energy schemes, solar photovoltaic (PV) and thermophotovoltaic (TPV) energy conversion report high efficiencies. In contrast to solar PVs that convert sunlight to electricity, a TPV system involves a hot thermal emitter that transforms heat to thermal radiation, which in turn is converted to electricity via a PV cell. The use of the thermal emitter in the conversion of heat to electricity opens a large parameter space for photonic engineering, and promises improved peformance.}

\par{The potential for efficiency improvement by use of TPV systems can be seen by comparing the efficiencies of solar PV systems with solar TPV systems. In the latter case, sunlight is used to provide the heat to the thermal emitter. Since both systems are photonic heat engines, their efficiencies can be compared to the Carnot efficiencly limit, $\eta_\mathrm{Carnot}=1-T_\mathrm{C}/T_\mathrm{H}$, where $T_\mathrm{H/C}$ is the temperature of the heat pump/sink of the heat engine. For solar energy conversion, $T_\mathrm{H}$ refers to the temperature of the sun, $T_\mathrm{Sun} \approx 6000$ K, and $T_\mathrm{C}$ refers to the temperature of the cell commonly taken to be $300$ K. For such a choice of $T_\mathrm{H}$ and $T_\mathrm{C}$, the Shockley-Queisser efficiency limit of a single junction solar cell is $30\%$ \cite{DetailedBalance1961}, whereas equivalent detailed balance analysis of a solar TPV system yields an efficiency limit of about $54\%$ \cite{Harder_2003}.}

\par{In addition to solar energy harvesting, TPV systems can also be used to harvest heat that is provided to the thermal emitter by other means. For example, a record-high TPV efficiency of $30\%$ was recently reported experimentally for a heat source at $T_\mathrm{H}=1455$ K \cite{Record2020_TPV_Forrest}. For both solar and thermal energy harvesting, the promising performance of TPV energy conversion systems arises from the ability to control the spatial as well as spectral characteristic of thermal radiation in the heat exchange between the thermal emitter and PV cell.}

\par{Despite the promising performance metrics of TPV systems, practical challenges in terms of materials and optoelectronic design remain to be resolved. In particular, a key challenge in the development of TPV systems is to recycle low-grade (low-$T_\mathrm{H}$) waste heat into electricity \cite{DOE2008, Zhao_NFTPV}. This requires a reduction of the thermal emitter temperature with respect to recently demonstrated TPV systems \cite{Record2020_TPV_Forrest, Yablonovich_PNASTPV2019,Lipson_TPV_experiment, Reddy_NFTPV2018,Noda_TPV_2019}, while maintaining high efficiency. Using TPV systems for low-grade heat harvesting also requires low-band gap PV cells that, however, are subject to large non-radiative recombination losses.  Consequently, these cells exhibit far-below-unity external luminescence efficiency. In the theory of solar PVs, it is well-understood that a good solar PV ought to be a good light-emitting diode \cite{Yablonovitch_Good_LED2012}, which emphasizes the necessity for high external luminescence efficiency. A high efficiency TPV system that would be of practical relevance is subject to the same requirement of having a cell with a large external luminescence efficiency.}

\par{Motivated by the significant oppotunties as well as challenges of TPV systems, in this paper, we carry out a detailed balance analysis of near-field TPV systems, where the spacing between the hot emitter and the cell is smaller than the thermal wavelength. In this range, the thermal energy transfer to the cell is dominated by near-field evanescent modes. Near-field TPV systems have been extensively considered in previous works for their potential to enhance the TPV current and therefore the extracted electrical power density, owing to an enhancement in the photonic thermal power density delivered to the cell as compared to far-field systems \cite{Park_NFTPV2008, Zhao_NFTPV, Zhao_thermophotonic2018, Papadakis_2019TPV, Francoeur_NFTPV2011,Francoeur_NFTPV2017}. In contrast to previous works, here, we focus on the physics of the open-circuit voltage. We show that a near-field TPV system can lead to significant enhancement of the external luminescent efficiency, and hence the open-circuit voltage, as compared with the far-field TPV systems and solar PV systems.}

\par{We identify two important opportunities in near-field TPV systems for the enhancement of the open-circuit voltage. First, in the near-field, the density of photon states that the cell can emit into is significantly enhanced as compared to the far-field. This enhancement can lead to a significant increase of the radiative recombination rate in the cell. Second, since non-radiative recombination is a volumetric effect, the use of a thinner cell leads to a reduction of the non-radiative recombination rate per unit area of the cell. Therefore, in a near-field TPV system with proper design, including the use of a high-reflectivity mirror behind the cell that re-directs photons that are not absorbed by the cell to the emitter where they are re-absorbed (Fig. \ref{fig:Figure1}), one can use a much thinner cell as compared to what is typically used in solar PV systems as well as what has been considered in previous TPV literature \cite{Greffet_NFTPV2006,Ilic_TPV2012,Zhao_NFTPV,Papadakis_2019TPV,Francoeur_NFTPV2011,Francoeur_NFTPV2017,Kaifeng_Cooling_2015}. We show that the combination of these effects leads to very significant enhancement of the open-circuit voltage. Since the open-circuit voltage is thermodynamically connected to the conversion efficiency, we show that thin-film near-field TPVs are advantageous in terms of conversion efficiency, in addition to electrical power density, as compared to their far-field counterparts.}

\par{A key innovation in our work is the development of an analytic model of a near-field TPV system. Unike research on solar PV systems, where there have been substantial efforts in developing insightful analytic frameworks \cite{PolmanAtwater_highefficiency, Rau_Voc_reciprocity_2007, Green_downconversion, Markvart_Solar_Heat_Engine, Rau_Photon_Recycling}, most existing works on near-field TPV energy conversion use numerical simulations based on fluctuational electrodynamics. While these simulations can accurately predict the performance of specific near-field TPV system designs, as was done in  \cite{Francoeur_NFTPV2017} where the interplay between radiative and non-radiative recombination was considered similar to the current work, it is difficult via numerical results to develop a more global understanding of the effects of various factors in controlling the performance of near-field TPV systems. Previous attempts to develop an analytic model utilize a highly idealized near-field blackbody model that is known to significantly overestimate the power density in near-field radiative heat transfer \cite{Pendry_NFHT1999}. In this paper, we introduce a new, simple analytic model that takes into account the narrowband nature of near-field radiative heat transfer between the emitter and the cell as well as the existence of a wavevector cutoff in the heat transfer \cite{Abdallah_betaSPP, Max_K_Biehs2010, Max_K_Francoer_2011}. Our model agrees very well with simulations based on fluctuational electrodynamics, and accurately describes performance metrics of near-field TPV systems.}

\par{The rest of the paper is organized as follows: In Section \ref{Theory}, we carry out a detailed balance analysis of the open-circuit voltage, which is discussed in terms of radiative and non-radiative processes taking place in the cell, discussed separately in Sections \ref{section:Vrad} and \ref{section:Vnrad}. In these sections we discuss our analytic theory of near-field TPV systems and compare to the standard treatment with fluctuational electrodynamics. Finally, in Section \ref{section:Performance} we carry out numerical calculations of an InAs-based thin-film near-field TPV system, where we show that the aforementioned opportunities associated with operating a TPV in the near-field can yield covnersion efficiencies $>50\%$ at practically relevant emitter temperatures.}

\section{\label{Theory}Theoretical Formalism}
 
 \subsection{\label{Detailbalance} Detailed Balance}
 
\par{We start by considering a hot emitter at a temperature $T_\mathrm{H}$ that faces a PV cell. The cell has a band gap of energy $\hbar\omega_\mathrm{g}$ and is maintained at a temperature $T_\mathrm{C}$. The emitter could be either a thermally emitting material in a TPV system or the sun in a solar PV system. Based on the principle of detailed balance, the current density in the cell, $J(V)$, is expressed as \cite{DetailedBalance1961}:
\begin{equation}\label{eq:1}
J(V)=J_\mathrm{e}-J_\mathrm{o}e^{qV/kT_\mathrm{C}}+R_\mathrm{o}-R(V),
\end{equation}
where $q$ is the electron charge, $k$ is the Boltzmann constant and $V$ is the voltage of the cell. The term $J_\mathrm{e}$ expresses the radiative generation of electron hole pairs from the influx of photons from the  emitter, referred to as the pump current henceforth, whereas $R_\mathrm{o}$ is the non-radiative generation of electron-hole pairs per unit area in the cell. The term $J_\mathrm{o}$ is the current density generated from radiative recombination (spontaneous emission) in the cell, or luminescence current, whereas $R(V)$ is the non-radiative recombination current. Following Shockley-Queisser analysis \cite{DetailedBalance1961}, we approximate $R(V)=R_\mathrm{o}e^{qV/kT_\mathrm{C}}$, a relation that we shall revisit later (Section \ref{section:Performance}). By solving for $J(V_\mathrm{oc})=0$ we obtain the open-circuit voltage:
\begin{equation}\label{eq:2}
qV_\mathrm{oc}=kT_\mathrm{C}\mathrm{ln}(\frac{J_\mathrm{e}}{J_\mathrm{o}})+kT_\mathrm{C}\mathrm{ln}(1+\frac{R_\mathrm{o}}{J_\mathrm{e}})+kT_\mathrm{C}\mathrm{ln}(\frac{J_\mathrm{o}}{J_\mathrm{o}+R_\mathrm{o}}).
\end{equation}
$qV_\mathrm{oc}$ is the maximum amount of electrical energy that could be extracted from the cell, per incident photon. Hence, computing $V_\mathrm{oc}$ is important for determining the thermal radiation-to-electricity conversion efficiency.}

\par{The first term in Eq. \ref{eq:2} depends solely on radiative processes, namely the absorption of thermal photons from the emitter and the radiative recombination of the cell. We define the radiative open-circuit voltage as:
\begin{equation}\label{eq:Vocrad}
qV_\mathrm{oc,rad}=kT_\mathrm{C}\mathrm{ln}(J_\mathrm{e}/J_\mathrm{o}).
\end{equation}
For a typical TPV system, the emitter temperature is sufficiently high such that $J_\mathrm{e} \gg R_\mathrm{o}$, therefore the second term in Eq. \ref{eq:2} is omitted henceforth. The third term in Eq. \ref{eq:2} depends on the interplay between radiative and non-radiative processes of the cell. By defining the external luminescence efficiency as:
\begin{equation}\label{eq:Qe}
Q_\mathrm{e}=J_\mathrm{o}/(J_\mathrm{o}+R_\mathrm{o}),
\end{equation}
we see that the presence of nonradiative recombination yields a negative contribution to the open circuit voltage, namely:
\begin{equation}\label{eq:Vocnrad}
V_\mathrm{oc,nrad}=kT_\mathrm{C}\mathrm{ln}(Q_\mathrm{e}).
\end{equation}
So far, we have separated the open-circuit voltage into a radiative term, $V_\mathrm{oc,rad}>0$, and a non-radiative one, $V_\mathrm{oc,nrad}<0$, and we can write:
\begin{equation}\label{eq:3}
qV_\mathrm{oc}=qV_\mathrm{oc,rad}+qV_\mathrm{oc,nrad}.
\end{equation}
In the following sections, we will analyze these contributions separately.}

\par{The pump current ($J_\mathrm{e}$) and luminescence current ($J_\mathrm{o}$) in Eqs. \ref{eq:1}, \ref{eq:2} are determined with fluctuational electrodynamics \cite{PolderNFHT} via:
\begin{equation}\label{eq:4}
J_\mathrm{o/e}=\frac{q}{4\pi^2}\int_{\omega_\mathrm{g}}^{\infty} \Phi_\mathrm{C/H}(\omega) n(\omega, T_\mathrm{C/H})d\omega,
\end{equation}
where $\omega$ is the angular frequency, $\Phi_\mathrm{C}(\omega)$ and $\Phi_\mathrm{H}(\omega)$ are the thermal emission spectra of the cell and the emitter, respectively, and $n(\omega,T)$ is the photon occupation number, which we approximate here by $n(\omega,T)=e^{-\hbar\omega/kT}$, assuming $\hbar\omega_\mathrm{g}\gg k T_\mathrm{C/H}$. By considering emission from a planar surface, we can write the spectra $\Phi_\mathrm{C/H}(\omega)$ as
\begin{equation}\label{eq:5}
\Phi_\mathrm{C/H}(\omega)=\int_{0}^{\beta_\mathrm{max,C/H}} \xi(\omega, \beta) \beta d\beta,
\end{equation}
where $\beta$ refers to the in-plane wavenumber of the modes participating in the radiative heat exchange between the emitter and the cell. The term $\xi$ in Eq. \ref{eq:5} is the probability of a photon emitted by body C/H with frequency $\omega$ and wavenumber $\beta$ to be absorbed by body H/C. The upper limit of integration in Eq. \ref{eq:5}, $\beta_\mathrm{max,C/H}$, refers to the maximum in-plane wavenumber that participates in the radiative heat exchange between the emitter and the cell. Below, we examine the the pump current ($J_\mathrm{e}$) and luminescence current ($J_\mathrm{o}$) for the cases of a solar PV, a far-field TPV system, and a near-field TPV system.}

\subsection{\label{SolarPV&FFTPV} Solar PV cell and far-field TPV cell}
 
 \begin{figure}[]
\centering
\includegraphics[width=1\linewidth]{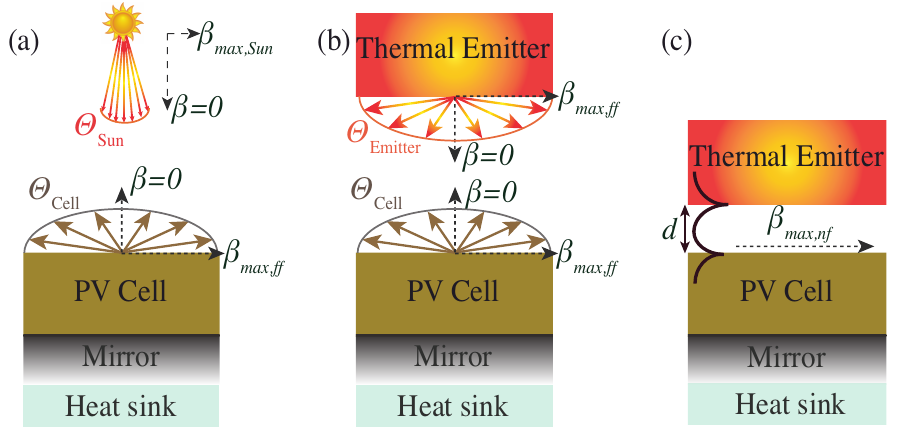}
  \caption{Schematics of (a) solar PV cell (without concentration), (b) a planar far-field TPV system, and (c) a near-field TPV system. The quantities $\Theta_\mathrm{Sun}$, $\Theta_\mathrm{Emitter}$, $\Theta_\mathrm{Cell}$ define the light cones of emission of each body, and $\beta_\mathrm{max}$ is the corresponding maximum in-plane wavenumber of the electromagnetic modes that participate in the delivery of thermal photons.}
 \label{fig:Figure1} 
\end{figure}

\par{The exchange of thermal radiation for a solar PV cell, shown schematically in Fig. \ref{fig:Figure1}(a), and for a far-field TPV system, shown in Fig. \ref{fig:Figure1}(b), occurs exclusively via propagating modes. In the case of a far-field TPV system, we assume that the size of the vacuum gap separating the emitter from the cell is much larger than the relevant wavelengths. We approximate the sun and thermal emitter in Figs. \ref{fig:Figure1}(a) and (b), respectively, as blackbodies, and assume that the cell has unity absorptivity above the band gap, and zero absorptivity below, hence we consider $\xi_\mathrm{ff}=1$ across all frequencies above the band gap. The subscript ``ff'' refers to far-field, pertaining to a cell that is placed in the far-field of a thermal emitter, or the sun.}

\par{In either case, concerning the luminescence current, $J_\mathrm{o}$, thermal radiation from the PV cell is emitted into a hemisphere, therefore the solid angle of emission is $2\pi$ steradians, corresponding to an apex angle of $\Theta_\mathrm{Cell}=\pi$. Hence, the maximum in-plane wavenumber is $\beta_\mathrm{max,C}=(\omega/c)\mathrm{sin}(\Theta_\mathrm{Cell}/2)=\omega/c$ where $c$ is the speed of light (see Fig. \ref{fig:Figure1}). From Eqs. \ref{eq:4}, \ref{eq:5}, the luminescence current of PV cell placed in the far-field of a thermal emitter or the sun is given by:
\begin{equation}\label{eq:6}
J_\mathrm{o,ff}=\frac{q}{4\pi^2}\frac{kT_\mathrm{C}\omega_\mathrm{g}^2}{2\hbar c^2}e^{-\hbar\omega_\mathrm{g}/kT_\mathrm{C}}.
\end{equation}
Eq. \ref{eq:6} connects the radiative recombination, in other words the spontaneous emission, of a far-field PV cell with its temperature and band gap, and will be used in the following sections in estimating its open-circuit voltage.}

\par{Regarding the pump current, $J_\mathrm{e}$, the cases of a solar PV and a far-field TPV are fundamentally different. First, we consider the case of solar thermal emission as shown in Fig. \ref{fig:Figure1}(a). The pump current coming from the sun as received by the PV cell, $J_\mathrm{e,Sun}$, is restricted by the narrow light cone of the sun. Particularly, the solid angle of emission of the sun as seen by an object on earth is $6.8 \times 10^{-5}$ steradians \cite{PolmanAtwater_highefficiency}, corresponding to an apex angle of $\Theta_\mathrm{Sun}=9.2\times 10^{-3}$ radians. Thereby, the maximum wavenumber that participates in the radiative heat transfer from the sun is $\beta_\mathrm{max,Sun}=(\omega/c)\mathrm{sin}(\Theta_\mathrm{Sun}/2)$, which is significantly smaller than $\omega/c$, therefore the integration in Eq. \ref{eq:5} covers a smaller angular range as compared to the emission from a PV cell (Eq. \ref{eq:6}). The pump current of a solar PV is: 
\begin{equation}\label{eq:10}
J_\mathrm{e,Sun}=\frac{q}{4\pi^2}\frac{kT_\mathrm{Sun}\omega_\mathrm{g}^2 e^{-\hbar\omega_\mathrm{g}/kT_\mathrm{Sun}}}{2\hbar c^2}\mathrm{sin}^2(\frac{\Theta_\mathrm{Sun}}{2}),
\end{equation}
where $T_\mathrm{Sun}=6000$ K. As we shall see in Section \ref{section:Vrad}, the last factor in Eq. \ref{eq:10} significantly restricts the radiative open-circuit voltage of solar PV cells.}

\par{In the case of a far-field TPV system, a planar thermal emitter (Fig. \ref{fig:Figure1}(b)) emits into a hemisphere, thereby the solid angle of emission is $2\pi$ steradians corresponding to an apex angle of $\Theta_\mathrm{Emitter}=\Theta_\mathrm{Cell}=\pi$. Hence, the emitter and cell emit into the same light cone, hence the view factor is unity, assuming that the surface area of the emitter  and the cell are large with respect to their separation. In this case, the integration of Eq. \ref{eq:5} covers the same angular range as in the case of $J_\mathrm{o,FF}$, i.e. $\beta_\mathrm{max,H}=\beta_\mathrm{max,C}=\beta_\mathrm{max,ff}=\omega/c$ (see Fig. \ref{fig:Figure1}(b)). The pump current is given by: 
\begin{equation}\label{eq:10b}
J_\mathrm{e,ff}=\frac{q}{4\pi^2}\frac{kT_\mathrm{H}\omega_\mathrm{g}^2 e^{-\hbar\omega_\mathrm{g}/kT_\mathrm{H}}}{2\hbar c^2}.
\end{equation}
Eqs. \ref{eq:6}, \ref{eq:10}, \ref{eq:10b} will allow us to estimate the radiative open-circuit voltage, as defined in Eq. \ref{eq:Vocrad}, in Section \ref{section:Vnrad}.}

 \subsection{\label{NFTPV} Near-field TPV cell}

\par{The case of a near-field TPV system is shown schematically in Fig. \ref{fig:Figure1}(c), where the size of the vacuum gap, $d$, is assumed to be comparable to or smaller than the wavelength, while also being significantly smaller than the lateral dimensions of the planar emitter and cell. In this case, thermal radiation between the emitter and the cell is exchanged via both propagating and evanescent modes. For the evanescent modes, $\xi$ in Eq. \ref{eq:5} becomes the probability of a photon to tunnel through the sub-wavelength vacuum gap, summed over the two polarizations. The near-field provides access to wavenumbers significantly larger than $\omega/c$. Theoretically, the integration in Eq. \ref{eq:5} is over a range of $\beta$'s extending to $\infty$ for the case of near-field heat transfer via evanescent modes. In practice, however, there is a maximum wavenumber beyond which the contribution of the integrand becomes negligible. Similar to the case of far-field TPV systems, due to a view factor of unity, $\beta_\mathrm{max,C}=\beta_\mathrm{max,H}=\beta_\mathrm{max,nf}$ (see Fig. \ref{fig:Figure1}(c)), where the subscript ``nf'' corresponds to near-field. An accurate description of $\beta_\mathrm{max,nf}$ is critical in estimating the performance of realistic near-field TPV systems.}

\par{It is broadly considered that the maximum wavenumber for near-field heat transfer is $\beta_\mathrm{max,nf}\approx C/d$, where $C$ is a constant with a magnitude on the order of unity \cite{Pendry_NFHT1999,Abdallah_betaSPP,Max_K_Biehs2010,Francoeur_NFTPV2017,Max_K_Francoer_2011}. Since the vacuum gap is on the order of hundreds to tens of nanometers in the near-field, $\beta_\mathrm{max,nf}$ obtains very large values compared to the far-field.}
\par{As a simple model of near-field heat transfer \cite{Pendry_NFHT1999}, one assumes that the photon tunneling probability vanishes at frequencies below the band gap whereas it maximizes above the band gap. Namely, 
\begin{equation}\label{eq:blackbodyNF}
\xi_\mathrm{BB}(\omega,\beta) =
    \begin{cases}
      0 & \text{for $\omega<\omega_\mathrm{g}$, all $\beta$'s}\\
      1 & \text{for $\omega\geq\omega_\mathrm{g}$ and $\beta\leq C/d$}\\
      0 & \text{for $\omega\geq\omega_\mathrm{g}$ and $\beta>C/d$}.
    \end{cases}
  \end{equation}
The case $\xi=1$ pertains to perfect thermal emission, and by reciprocity, perfect absorption, in the near-field. The equivalent properties in the far-field correspond to a perfect blackbody. Even though the notion of a blackbody generally refers to far-field thermal radiation, in what follows, we shall refer to this case as the blackbody model, and index relevant parameters with ``BB'' as in Eqs. \ref{eq:blackbodyNF}, \ref{eq:7}. For $C=1$ as in \cite{Pendry_NFHT1999}, from Eqs. \ref{eq:4}, \ref{eq:5} one can derive the luminescence current in the near-field:
\begin{equation}\label{eq:7}
J_\mathrm{o,nf,BB}=\frac{q}{4\pi^2}\frac{kT_\mathrm{C}}{2\hbar d^2}e^{-\hbar\omega_\mathrm{g}/kT_\mathrm{C}}.
\end{equation}
}

\par
\begin{figure}[]
\centering
\includegraphics[width=1\linewidth]{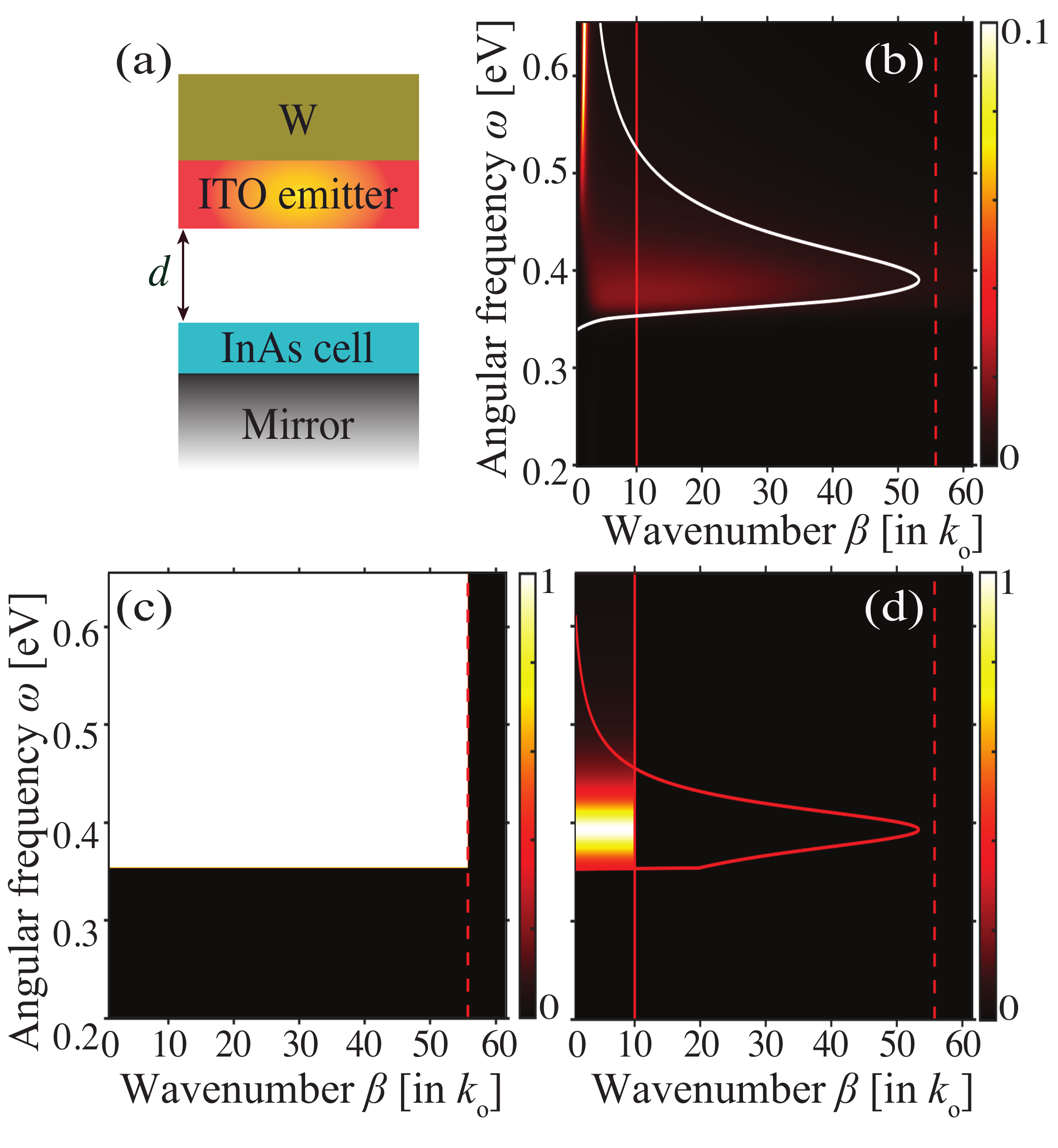}
   \caption{(a) Schematic of a near-field TPV system composed of an InAs cell of thickness $20$ nm and an ITO thermal emitter with $\omega_\mathrm{p}=0.4$ eV, such that it matches the band gap of InAs. (b), (c), (d) Contour plots of the photon tunneling probability, $\xi(\omega,\beta)$, obtained via fluctuational electrodynamics-(b), the blackbody model-(c), and the narrowband model (d), for $d=10$ nm. The dashed and solid red vertical lines show the maximum in-plane wavenumber for the blackbody model ($1/d$) and the narrowband model ($\beta_\mathrm{max,NR}=0.18/d$), respectively. The solid white and red curves in panels (b) and (d) are the spectra $\Phi(\omega)$ obtained via fluctuational electrodynamics and via the narrowband model, respectively, in arbitrary units.}
 \label{fig:Figure2} 
\end{figure}

{Unfortunately, the blackbody model does not accurately describe the heat transfer spectrum in typical near-field TPV systems. To illustrate this, as a typical near-field TPV case, we consider a PV cell coupled to a thermal emitter that supports a plasmonic mode, as has been the case in recent literature in near-field TPVs \cite{Greffet_NFTPV2006, Zhao_NFTPV, Ideal_NFTPV_Jabob, Papadakis_2019TPV, Ilic_TPV2012,Francoeur_NFTPV2017, Joannopoulos_SqueezingNFHT}. Here, similar to \cite{Zhao_NFTPV, Papadakis_2019TPV}, we consider an InAs PV cell on a perfectly reflecting back-side mirror for photon recycling purposes \cite{Yablonovich_PNASTPV2019} as shown in Fig. \ref{fig:Figure2}(a). The thermal emitter is composed of a $30$ nm indium tin oxide (ITO) film on a tungsten (W) back-side heat spreader. The band gap of InAs is $\hbar\omega_\mathrm{g,InAs}=0.354$ eV and the plasma frequency of the ITO emitter is set to $0.4$ eV in units of energy. We note that, in contrast to previous considerations of near-field TPV systems including \cite{Zhao_NFTPV, Papadakis_2019TPV}, here, we consider an ultra-thin InAs cell of $20$ nm for enhancing the conversion efficiency, as discussed in more detail in Section \ref{section:Vnrad}.}

\par{As a benchmark for evaluating the accuracy of the blackbody model introduced above, we perform exact calculations of near-field heat transfer via fluctuational electrodynamics, following the original formalism in \cite{PolderNFHT} that is based on Fresnel's coefficients for layered media, computed here via \cite{CHEN2018163}. In Fig. \ref{fig:Figure2}(b), we display the photon tunneling probability for the considered near-field TPV system at a vacuum gap of $d=10$ nm as obtained with fluctuational electrodynamics, and it can be seen that it hardly reaches the value of $0.1$ for a narrow frequency range near the band gap of InAs ($0.354$ eV). In contrast, in Fig. \ref{fig:Figure2}(c) we show the blackbody model (Eq. \ref{eq:blackbodyNF}), which asserts a photon tunneling probability of unity across all frequencies above the band gap and across all wavenumbers up to $1/d$, shown with the red dashed vertical line in Figs. \ref{fig:Figure2}(b)-(d). It therefore becomes apparent that, although the blackbody model is useful in predicting an upper bound to the near-field heat transfer \cite{Pendry_NFHT1999}, it is highly idealized and severely deviates from a realistic situation. No known material systems exhibit the broadband near-unity emissivity in the near-field as described by the blackbody model. Therefore, a model that accurately captures the spectral characteristics of near-field radiative heat transfer in TPV systems is necessary for estimating performance metrics like open-circuit voltage, current density, and efficiency.}

\par{In introducing such a model, we consider that near-field radiative heat transfer is significantly more narrowband in nature as compared to far-field thermal radiation, as has been shown previously \cite{GangChen_NFHTmeasurement2008, Raschke_NFHTReview, NFTPV_review2008, Abdallah_ReviewNFTPV,Greffet_NFTPV2006,Papadakis_2019TPV}. Thereby, we introduce the following phenomenological photon tunneling probability based on a Lorentzian lineshape:
\begin{equation}\label{eq:8}
\xi_\mathrm{NR}(\omega,\beta)=
 \begin{cases}
      0 & \text{for $\omega<\omega_\mathrm{g}$, all $\beta$'s}\\
      \frac{\Gamma^2}{\Gamma^2+(\omega-\omega_\mathrm{o})^2}  & \text{for $\omega\geq\omega_\mathrm{g}$ and $\beta\leq \rho/d$}\\
      0 & \text{for $\omega\geq\omega_\mathrm{g}$ and $\beta>\rho/d$},
    \end{cases}
\end{equation}
 where the index ``NR'' refers to narrowband near-field emission. The parameter $\Gamma$ is the bandwidth of the near-field interaction, and $\omega_\mathrm{o}$ is slightly smaller than $\omega_\mathrm{g}$. Therefore, from Eq. \ref{eq:8}, as $\omega$ approaches the band gap, the photon tunneling probability increases, whereas away from the band gap it rapidly decreases to values far below unity, as is the case for realistic material systems.}
 
 \par{In the model of Eq. \ref{eq:8}, $\rho\ll1$, hence the maximum wavenumber participating in near-field heat transfer is now scaled by this quantity, i.e. $\beta_\mathrm{max,NR}=\rho/d$. This captures the fact that that the maximum in-plane wavenumber in near-field heat transfer between realistic material systems is significantly smaller than $1/d$. Furthermore, it also reduces as the thickness of the materials decreases. The former has been previously discussed for the case of near-field heat transfer between identical polaritonic materials \cite{Abdallah_betaSPP, Papadakis_TunableNFHT,Max_K_Biehs2010}, nevertheless it has not been considered within the context of near-field TPV systems, where the heat exchange occurs typically between a semiconductor and a polaritonic material. The quantity $\rho$ can be formally obtained via:
 \begin{equation}\label{eq:rho}
\int_{0}^{\rho/d} \xi_\mathrm{NR}(\omega, \beta) \beta d\beta=\int_{0}^{\infty} \xi(\omega, \beta) \beta d\beta,
\end{equation}
where $\xi(\omega,\beta)$ is the exact photon tunneling probability as obtained with fluctuational electrodynamics. This definition of $\rho$ ensures the agreement between the spectral characteristics of thermal radiation obtained via the narrowband model and those obtained via fluctuational electrodynamics. Due to the narrowband nature of near-field heat transfer, setting $\omega=\omega_\mathrm{g}$ in Eq. \ref{eq:rho}, suffices. In practice, $\beta_\mathrm{max,NR}$ can be approximated by searching for the value of $\beta$ for which $\xi(\omega,\beta)$, obtained via fluctuational electrodynamics using, for example, Fresnel's equations or a computational package like \cite{CHEN2018163}, is maximized.}

\par{In Fig. \ref{fig:Figure2}(d), we display the photon tunneling probability as obtained with the narrowband model for the InAs/ITO TPV system introduced above. The bandwidth $\Gamma$ is set to $30$ meV, which corresponds to $3\gamma_\mathrm{ITO}$, where $\gamma_\mathrm{ITO}$ is the loss rate in the Drude model used to describe the dielectric function of the ITO thermal emitter \cite{Papadakis_2019TPV}. Furthermore, $\rho=0.18$ hence $\beta_\mathrm{max} \sim 0.18/d$. This wavenumber is shown with the solid red vertical line in the contour plots of Figs. \ref{fig:Figure2}(b)-(d) and it is significantly smaller than $1/d$. The photon tunneling probability of the narrowband model does \textit{not} agree with fluctuational electrodynamics (Fig. \ref{fig:Figure2}(b)). Nevertheless, as discussed above, the construction of the narrowband model ensures that the spectral characteristics of thermal emission obtained with the narrowband model closely resemble those obtained with fluctuational electrodynamics. This can be seen by comparing the white and red curves in panels (b) and (d) of Fig. \ref{fig:Figure2}, obtained with fluctuational electrodynamics and the narrowband model, respectively.}

\par{To complete this section, we compute the luminescence current for the narrowband model (Eq. \ref{eq:8}). From Eqs. \ref{eq:4}, \ref{eq:5}, for $\hbar\Gamma\ll kT_\mathrm{C}$ this is (see Supplemental Information):
\begin{equation}\label{eq:9}
J_\mathrm{o,nf,NR}=\frac{q}{4\pi}\frac{\rho^2 \Gamma}{2 d^2}e^{-\hbar\omega_\mathrm{g}/kT_\mathrm{C}}.
\end{equation}
From Eq. \ref{eq:9}, the near-field luminescence current increases with both the bandwidth of the near-field thermal radiation, $\Gamma$, as well as the parameter $\rho$ that controls the cutoff value of the  maximum wavenumber, as expected.}

\section{Radiative $V_\mathrm{oc,rad}$}\label{section:Vrad}

\par{We evaluate the term $V_\mathrm{oc,rad}$ of Eq. \ref{eq:Vocrad} for the case of near-field TPV systems, and compare it with that of solar PV systems and far-field TPV systems, as introduced above.}\\

\noindent{\textbf{Solar PVs.} The case of solar PVs is shown schematically in Fig. \ref{fig:Figure1}(a), whereas the luminescence current and pump current from the sun are given, respectively, from Eqs. \ref{eq:6}, \ref{eq:10}. Hence, the radiative open-circuit voltage of a solar PV is:
\begin{equation}\label{eq:11}
    \begin{aligned}
qV_\mathrm{oc,solar,rad}=\hbar\omega_\mathrm{g}(1-\frac{T_\mathrm{C}}{T_\mathrm{Sun}})+kT_\mathrm{C}\mathrm{ln}(\frac{T_\mathrm{Sun}}{T_\mathrm{C}})\\
+2kT_\mathrm{C}\mathrm{ln}(\mathrm{sin}\frac{\Theta_\mathrm{Sun}}{2}).
  \end{aligned}
\end{equation}}
\par{The last term in Eq. \ref{eq:11} reaches nearly $-0.35$ eV, corresponding to over $25\%$ of the band gap of standard solar PVs (e.g. Si, GaAs).}\\

\noindent{\textbf{Far-field TPVs.} In the case of a far-field TPV system as the one shown in Fig. \ref{fig:Figure1}(b), the luminescence current remains that of Eq. \ref{eq:6}, whereas the pump current is given by Eq. \ref{eq:10b}. Therefore, the radiative open-circuit voltage becomes:
\begin{equation}\label{eq:12}
qV_\mathrm{oc,BB,rad}=\hbar\omega_\mathrm{g}(1-\frac{T_\mathrm{C}}{T_\mathrm{H}})+kT_\mathrm{C}\mathrm{ln}(\frac{T_\mathrm{H}}{T_\mathrm{C}}).
\end{equation}
By comparing Eqs. \ref{eq:11}, \ref{eq:12}, it becomes apparent that unconcentrated solar PV systems are limited by an additional angle-corrected loss mechanism, represented by the last term in Eq. \ref{eq:11}, which is absent in TPV systems (Eq. \ref{eq:12}). This term originates from the narrow light cone of the sun as compared to that of the cell. To mitigate these losses in solar PV systems, a standard approach is the use of solar concentrators \cite{WINSTON1975}, where parabolic mirrors are employed to increase the light cone of the sun that the cell sees, and therefore to increase $\Theta_\mathrm{Sun}$. This increases the solar pump current, $J_\mathrm{e,Sun}$, and therefore enhances $V_\mathrm{oc,rad}$ (See Eq. \ref{eq:Vocrad}). Alternatively, angle-selection approaches have been proposed \cite{Kosten_Angle_Restriction,PolmanAtwater_highefficiency} that aim to significantly narrow the light cone of the PV cell such that it matches the narrow light cone of the sun by decreasing $\Theta_\mathrm{Cell}$. In this case, the luminescence current is reduced, which increases $V_\mathrm{oc,rad}$ as seen by Eq. \ref{eq:Vocrad}. However, in the presence of non-radiative recombination, this approach additionally limits the luminescence efficiency (Eq. \ref{eq:Qe}). In contrast, in the case of TPV systems, the planarity of the emitter and cell warrant a view factor of unity, resulting in perfect matching between the light cone of the emitter and cell. Furthermore, from Eq. \ref{eq:12}, we see that that the open-circuit voltage of a TPV system can exceed the product $\hbar \omega_\mathrm{g} \eta_\mathrm{Carnot}$, which is often regarded as an upper bound to $V_\mathrm{oc}$ \cite{PolmanAtwater_highefficiency,Ilic_TPV2012}.}\\

\noindent{\textbf{Near-field TPVs.} The case of a near-field TPV system in shown in Fig. \ref{fig:Figure1}(c). We will evaluate $V_\mathrm{oc,rad}$ with the blackbody model (Eqs. \ref{eq:blackbodyNF}, \ref{eq:7}) as well as with the narrowband model (Eqs. \ref{eq:8}, \ref{eq:9}). As mentioned previously, $\beta_\mathrm{max,H}=\beta_\mathrm{max,C}$, hence, with either model, the pump current can be obtained by exchanging $T_\mathrm{C}$ with $T_\mathrm{H}$ in Eqs. \ref{eq:7}, \ref{eq:9}, respectively.}

\par{With the blackbody model (Eq. \ref{eq:7}), it is straightforward to see that the radiative open-circuit voltage of a near-field TPV is identical to that derived in Eq. \ref{eq:12} for a far-field TPV system. Thus, as discussed above, the benefit of far-field TPV systems in terms of $V_\mathrm{oc,rad}$ also applies to near-field TPV systems, as described by the blackbody model.}

\par{By considering the narrowband near-field model and exchanging $T_\mathrm{C}$ with $T_\mathrm{H}$ in Eq. \ref{eq:9} to obtain the pump current, the radiative open-circuit voltage is:
\begin{equation}\label{eq:13}
qV_\mathrm{oc,NR,rad}=\hbar\omega_\mathrm{g}(1-\frac{T_\mathrm{C}}{T_\mathrm{H}}).
\end{equation}
We note that the result of Eq. \ref{eq:13} is in agreement with previous TPV literature in the limit of narrowband thermal emission \cite{Harder_2003,Papadakis_2019TPV}. Comparison between Eqs. \ref{eq:12} and \ref{eq:13} shows that the narrowband model predicts a slightly smaller $V_\mathrm{oc,rad}$ than the blackbody model. This reduction in $V_\mathrm{oc,rad}$ is expected, since $V_\mathrm{oc}$ reflects the amount of electrical energy extracted per incident photon. In the blackbody model, due to a broader bandwidth, the average energy per photon is higher as compared to the narrowband model.}

\par{In this section we computed the portion of the open-circuit voltage that pertains to radiative generation and recombination of charge carriers (Eq. \ref{eq:Vocrad}). We quantified an important difference in the entropic losses of solar PV systems as compared to TPV systems, arising from the mismatch between the light cone of a planar PV cell and the angular range of unconcentrated solar thermal radiation (see Eq. \ref{eq:11}). Due to a view factor of unity, such losses are absent in planar TPV systems.}
\par{Finally, we note that, in the absence of non-radiative losses, via Eq. \ref{eq:12}, operating in the near-field does \textit{not} provide an advantage in terms of $V_\mathrm{oc,rad}$, as compared to the far-field, thereby the radiative open-circuit voltage does not strongly depend on the vacuum gap thickness \cite{Francoeur_NFTPV2017}. By contrast, we shall see in the next section that there is a significant advantage in operating in the near-field for the open-circuit voltage, in the presence of non-radiative losses.}

\section{Non-radiative $V_\mathrm{oc,nrad}$ and luminescence enhancement}\label{section:Vnrad}

\par{In this section, we evaluate the term $V_\mathrm{oc,nrad}$ of Eq. \ref{eq:Vocnrad} for TPV systems in the presence of non-radiative recombination.}\

\par{From Eqs. \ref{eq:7}, \ref{eq:9}, both the blackbody model and the narrowband model predict that, in the near-field, the luminescence current, $J_\mathrm{o}$, increases inversely proportionally to $d^2$. This suggests that operating in the near-field can significantly enhance the external luminescence efficiency as defined in Eq. \ref{eq:Qe}, and hence the overall performance of the TPV system. In this section, we demonstrate this analytically as well as numerically, by studying the interplay between radiative and non-radiative recombination, i.e. terms $J_\mathrm{o}$ and $R_\mathrm{o}$ in Eq. \ref{eq:Qe}. Previous work in \cite{Francoeur_NFTPV2017} has considered the external luminescence efficiency for a specific material system. Furthermore, the enhancement of the luminescence current in the near-field with respect to non-radiative recombination has been discussed in \cite{Kaifeng_Cooling_2015}. In contrast to previous works, in what follows, we introduce a simple near-field enhancement parameter ($\alpha_\mathrm{NR}$) that allows for direct quantification of the $V_\mathrm{oc}$, pertaining to \textit{any} material system.}

\subsection{Analytical results}\label{section:Vnrad_analytic}

\par{We start by considering the case of the blackbody model as introduced in Sections \ref{SolarPV&FFTPV} and \ref{NFTPV}, pertaining to the far-field and near-field, as discussed in Eqs. \ref{eq:6}, \ref{eq:7}, respectively. From these equations, one can write $J_\mathrm{o,nf,BB}=\alpha_\mathrm{BB}^2 J_\mathrm{o,ff}$, where $\alpha_\mathrm{BB}=(1/2\pi)(\lambda_\mathrm{g}/d)$ expresses the enhancement of the luminescence current in the near-field, with $\lambda_\mathrm{g}$ being the band gap of the PV cell in units of wavelength. For moderate-temperature TPV systems, the band gap of the cell is smaller than $1$ eV, hence $\lambda_\mathrm{g}$ is on the order of few microns. Therefore, for a near-field TPV system with a vacuum gap thickness, $d$, on the order of tens of nm, the enhancement factor $\alpha_\mathrm{BB}$ is on the order of tens to hundreds. Hence, the near-field luminescence current is increased by a factor of tens of thousands with respect to the far-field.}

\par{One can express the near-field luminescence efficiency, $Q_\mathrm{e,nf,BB}$, in terms of the far-field luminescence current as:
\begin{equation}\label{eq:14}
Q_\mathrm{e,nf,BB}=\frac{J_\mathrm{o,ff}}{J_\mathrm{o,ff}+R_\mathrm{o}/\alpha_\mathrm{BB}^2}.
\end{equation}
From Eq. \ref{eq:14} we see that, by operating in the near-field, the effect that non-radiative recombination has on the external luminescence is effectively suppressed by $\alpha_\mathrm{BB}^2$.}

\par{As outlined in Section \ref{NFTPV}, the assumptions considered in the blackbody model are largely unrealistic. Hence, we consider next the narrowband model which, as discussed with respect to Fig. \ref{fig:Figure2}(d), accurately captures the spectral characteristics of near-field heat transfer in near-field TPV systems. Via Eq. \ref{eq:9} and Eq. \ref{eq:6}, we obtain the same scaling law for the luminescence efficiency as in Eq. \ref{eq:14}, however, with the narrowband model, the near-field enhancement factor $\alpha_\mathrm{NR}$ is given by:
\begin{equation}\label{eq:15}
\alpha_\mathrm{NR}^2=\rho^2 (\frac{\hbar\Gamma}{kT_\mathrm{C}}) \alpha_\mathrm{BB}^2.
\end{equation}
Since $\rho\ll1$, and since $\hbar\Gamma$ is typically comparable or smaller than $kT_\mathrm{C}$, the narrowband model predicts an enhancement factor that is considerably reduced compared to the one predicted with the blackbody model, i.e. $\alpha_\mathrm{NR}<\alpha_\mathrm{BB}$. Thereby, the narrowband model predicts reduced external luminescence efficiency with respect to the blackbody model.}

\par{We note that, in solar PV systems, large external luminescence is seen for very few semiconductors, for example GaAs-based heterostructures \cite{Yablonovitch_GaAs1992} and, recently, low-dimensional materials \cite{Javey_TMDs_2016}. In contrast, in the case of near-field TPV systems, the enhancement in luminescence efficiency is a natural consequence of operating in the near-field and arises from significant increase of spontaneous emission, i.e. the radiative recombination term $J_\mathrm{o}$ in Eq. \ref{eq:Qe}. Such enhancement is applicable for any PV cell material.}

\subsection{Numerical results}\label{section:Vnrad_numerical}

\par
\begin{figure}[]
\centering
\includegraphics[width=0.9\linewidth]{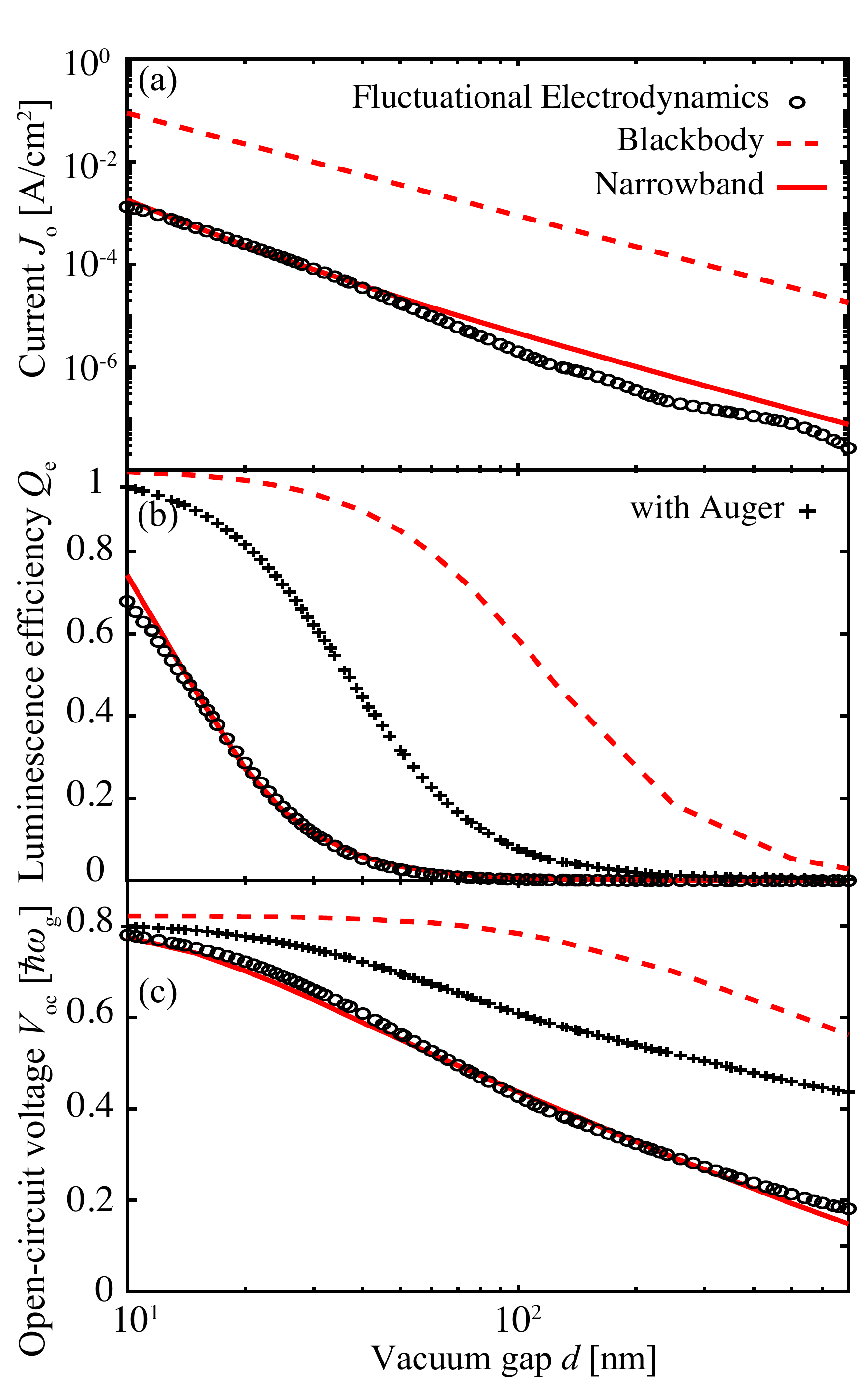}
   \caption{(a) Luminescence current, $J_\mathrm{o}$, (b) luminescence efficiency, $Q_\mathrm{e}$, and (c) open-circuit voltage, $V_\mathrm{oc}$, as a function of the vacuum gap size, $d$, for a thin-film near-field TPV composed of a $20$ nm cell with the optical properties of InAs and an ITO plasmonic emitter. These calculations correspond to the non-radiative recombination model $R=R_\mathrm{o,InAs}e^{qV/kT_\mathrm{C}}$ as described via detailed balance (Eqs. \ref{eq:1}, \ref{eq:2}), with $R_\mathrm{o,InAs}=6.3$ mA/$\mathrm{cm}^2$, for an emitter temperature of $T_\mathrm{H}=1100$ K. The red dashed curves show the blackbody model whereas the solid red curves pertain to the narrowband model, while the black circles pertain to fluctuational electrodynamics. The black crosses in panels (b), (c) corresponds to the realistic non-radiative model of Eq. \ref{eq:17}.}
 \label{fig:Figure3} 
\end{figure}

\begin{figure}[]
\centering
\includegraphics[width=0.9\linewidth]{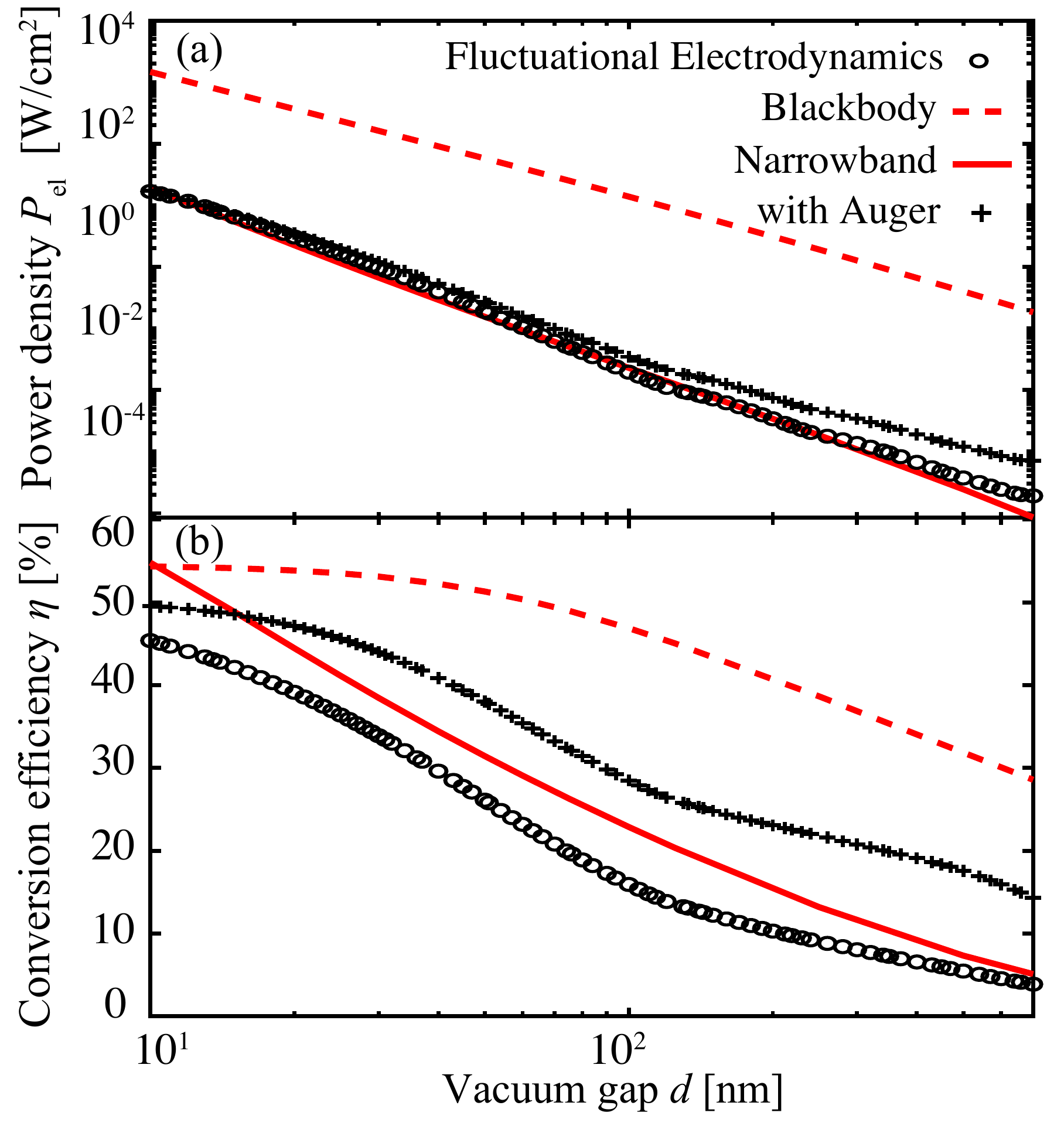}
   \caption{(a) Extracted electrical power density, $P_\mathrm{el}$, and (b) conversion efficiency, $\eta$, for the thin-film near-field TPV system discussed in Fig. \ref{fig:Figure3}, as a function of the vacuum gap size, $d$. The red dashed curves show the blackbody model whereas the red solid curves pertain to the narrowband model, while the black circles pertain to fluctuational electrodynamics. The black crosses pertain to fluctuational electrodynamics considering Auger non-radiative recombination (Eq. \ref{eq:17}).}
    \label{fig:Figure4} 
\end{figure}

{To demonstrate the results of the theory presented above, as discussed previously, we consider the near-field TPV system of Fig. \ref{fig:Figure2}(a). This system has been initially introduced in \cite{Zhao_NFTPV}, where the InAs cell was $400$ nm thick such that the extracted electrical power density is maximized. Here, in contrast, we consider a much thinner cell with thickness of $t_\mathrm{c}=20$ nm. Low-band gap semiconductors that are relevant for low-grade waste heat recycling via TPV energy conversion, such as InAs, are subject to significant non-radiative recombination, due to Auger and Shockley-Read-Hall (SRH) non-radiative processes \cite{Nonradiative_InAs}. These effects are volumetric (Eq. \ref{eq:17}), therefore decreasing the thickness of the cell suppresses their impact on the TPV performance. As we shall show, the InAs cell thickness reduction with respect to \cite{Zhao_NFTPV} yields a significant improvement in external luminescence efficiency and, consequently, in open-circuit voltage and conversion efficiency (Figs. \ref{fig:Figure3}, \ref{fig:Figure4}). Furthermore, we note that the use of a back-side mirror in the TPV design, as shown in Figs. \ref{fig:Figure1}(b), (c), \ref{fig:Figure2}(a), recycles photons that are not absorbed by the cell during their initial passage. A broadband mirror will recycle both below- and above-band gap photons, as has been recently shown experimentally in \cite{Yablonovich_PNASTPV2019, Record2020_TPV_Forrest}, thereby maximizing absorption and minimizing radiation leakage. Therefore, the reduction in cell thickness that we consider here does not yield a significant penalty in the extracted electrical power density, as we show below.}

\par{The dependence of Auger and SRH non-radiative recombination on the voltage, $V$, deviates from the non-radiative model we considered in Section \ref{Theory}, namely $R(V)=R_\mathrm{o}e^{qV/kT_\mathrm{C}}$ \cite{DetailedBalance1961} (see Eqs. \ref{eq:1}, \ref{eq:2}). Nevertheless, here, for the sake of a meaningful comparison between the blackbody model and narrowband near-field model, both of which consider $\sim e^{qV/kT_\mathrm{C}}$ dependence of non-radiative recombination, we will assume that the non-radiative losses in InAs obey the law $R(V)=R_\mathrm{o}e^{qV/kT_\mathrm{C}}$. In Figs. \ref{fig:Figure3}, \ref{fig:Figure4}, we show numerical results where we consider $R_\mathrm{o,InAs}=6.3$ mA/$\mathrm{cm}^2$. This value corresponds to the losses due to Auger non-radiative recombination for a cell thickness of $t_\mathrm{c}=20$ nm and n-type doping level of $2 \times 10^{16}$ $\mathrm{cm}^{-3}$, at a voltage of $qV=\hbar\omega_\mathrm{g,InAs}$. Since the open-circuit voltage is smaller than the band gap (Eqs. \ref{eq:12}, \ref{eq:13}), this value of $R_\mathrm{o,InAs}$ is an overestimation of the expected losses due to Auger recombination. In Section \ref{section:Performance}, we will consider the realistic non-radiative recombination model for InAs, for which results are shown in Figs. \ref{fig:Figure3}, \ref{fig:Figure4} in black.}

\par{In Figs. \ref{fig:Figure3}, \ref{fig:Figure4}, we compare the numerical results with the two near-field TPV models as introduced in previous sections. The numerical results are obtained through Eqs. \ref{eq:4} and \ref{eq:5}, where we carry out the integration using $\xi(\omega,\beta)$ as obtained from fluctuational electrodynamics calculations. For the blackbody model, the luminescence current is obtained via Eq. \ref{eq:6}.}

\par{To apply the narrowband model, we first compute the scaling parameter $\rho$ at each vacuum gap following the same procedure as discussed in Fig. \ref{fig:Figure2}(d). We find that the scaling law $\rho(d)=\rho_\mathrm{1}+\rho_\mathrm{2}/d$, with $\rho_\mathrm{1}=0.08$ and $\rho_\mathrm{2}=10^{-9}$ nm, where $d$ is the size of the vacuum gap in nm, accurately describes the result obtained with fluctuational electrodynamics. This dependence on the vacuum gap indicates that the near-field heat transfer between a semiconductor and a plasmonic material decreases more dramatically as $d$ increases, as compared to the predictions acquired with the blackbody model (where $\rho(d)=C=1$) \cite{Pendry_NFHT1999}. Therefore, the narrowband model predicts a significantly smaller maximum wavenumber for the whole range of vacuum gaps considered, as compared to the blackbody model. Based on this scaling law, we obtain the luminescence current directly from Eq. \ref{eq:9} for the narrowband model.}

\par{In Fig. \ref{fig:Figure3} we plot the luminescent current, $J_\mathrm{o}$ (Fig. \ref{fig:Figure3}(a)), the luminescent efficiency, $Q_\mathrm{e}$(Fig. \ref{fig:Figure3}(b)), and the open circuit voltage, $V_\mathrm{oc}$ (Fig. \ref{fig:Figure3}(c)), as a function of vacuum gap size, with both the numerical calculations using fluctuational electrodynamics (black circles) as well as the blackbody (red dashed curves) and the narrowband (red solid curves) models for near-field heat transfer. For all these quantities, we observe that the narrowband model is in excellent agreement with the results from fluctuational electrodynamics, whereas the blackbody model significantly overestimates. This is expected since the blackbody model severely overestimates the power density in near-field heat transfer, whereas the narrowband model accurately captures the spectrum of the heat exchange between the emitter and cell. Small deviations between the result with fluctuational electrodynamics and the narrowband model in the luminescent current (Fig. \ref{fig:Figure3}(a)) occur as $d$ increases, since the narrowband model pertains strictly to the near-field, whereas, as $d$ increases, propagating modes start to play an important role in the heat transfer mechanism.}

\par{In Fig. \ref{fig:Figure3}(a), we see that the luminescence current increases by more than four orders of magnitude as the gap size reduces from $0.7$ microns to $10$ nm. Such an increase is due to the enhancement of near-field heat transfer between the emitter and the cell and is consistent with previous calculations and experiments on near field heat transfer \cite{PolderNFHT,NFTPV_review2008,Raschke_NFHTReview,GangChen_NFHTmeasurement2008,GiantEnhancementSiO2_Reddy2018}. In a near-field TPV system, such enhancement in heat transfer results in a significant enhancement of the external luminescent efficiency, from $Q_\mathrm{e}=4 \times 10^{-5}$ at $d=0.7$ microns to $Q_\mathrm{e}=0.54$ at $d=10$ nm (Fig. \ref{fig:Figure3}(b)). As a result, the open circuit increases dramatically, from $V_\mathrm{oc}=0.048$ eV at $d=0.7$ microns to $V_\mathrm{oc}=0.27$ eV at $d=10$ nm (Fig. \ref{fig:Figure3}(c)). All these results are well accounted for by the analytic narrowband model.}

\par{In Fig. \ref{fig:Figure4} we show the maximum extracted electrical power density, $P_\mathrm{el}=J(V)\times V$ (Fig. \ref{fig:Figure4}(a)), and the conversion efficiency, $\eta$ (Fig. \ref{fig:Figure4}(b)), as a function of the vacuum gap. The conversion efficiency of a TPV is defined as $\eta=P_\mathrm{el}/P_\mathrm{phot}$, where $P_\mathrm{phot}$ is the photonic heat exchange between the emitter and the cell, expressed by: $P_\mathrm{phot}=P_\mathrm{\omega<\omega_\mathrm{g}}+P_\mathrm{\omega>\omega_\mathrm{g}}$, where the terms $P_\mathrm{\omega<\omega_\mathrm{g}}$ and $P_\mathrm{\omega>\omega_\mathrm{g}}$ correspond, respectively, to heat exchange below- and above-band gap. The above-band gap heat exchange is:
\begin{equation}\label{eq:16extra}
  P_\mathrm{\omega>\omega_\mathrm{g}}=P_\mathrm{e}-P_\mathrm{o}e^{qV/kT_\mathrm{C}},
\end{equation}
where $P_\mathrm{e}$ and $P_\mathrm{o}$ are the power densities of photon flux emitted by the emitter and cell, respectively. These are expressed as:
\begin{equation}\label{eq:16}
P_\mathrm{o/e}=\frac{1}{4\pi^2}\int_{\omega_\mathrm{g}}^{\infty} \hbar\omega\Phi_\mathrm{C/H}(\omega) n(\omega, T_\mathrm{C/H})d\omega,
\end{equation}
where $\Phi_\mathrm{C/H}$ were defined in Eq. \ref{eq:5}. The below-band gap heat exchange is obtained in a similar manner to Eq. \ref{eq:16extra}, however the voltage $V$ is set to zero and the integration range in Eq. \ref{eq:16} pertains to the frequencies below-band gap. In InAs, contributions to below-band gap near-field heat transfer originate largely from a surface phonon polariton mode that occurs at the Reststrahlen band at roughly $30$ meV \cite{Chen_FanNFTPV2015,Zhao_NFTPV, Papadakis_2019TPV}. Such contributions are included in the numerical results, but they are not captured by the narrowband or the blackbody models.}

\par{From Fig. \ref{fig:Figure4}, the narrowband model agrees well with the results from numerical calculations, whereas the blackbody model significantly overestimates the TPV performance metrics. As can be seen in Fig. \ref{fig:Figure4}(b), there is a small discrepancy in the efficiency between the result obtained with fluctuational electrodynamics (black dashed curve) and the narrowband model (blue curve). This discrepancy arises from the surface phonon polariton-mediated below-band gap heat exchange that is omitted in the narrowband model. As the vacuum gap size increases, this parasitic below-band gap heat transfer mechanism is suppressed, since surface-phonon polaritons are evanescent and decay rapidly as the vacuum gap increases \cite{Kfaifeng_Auger}. Hence, the agreement between the narrowband model and fluctuational electrodynamics improves as the vacuum gap increases.}

\par{From Fig. \ref{fig:Figure4}, as the vacuum gap size decreases, both the power density and the efficiency increase. Namely, as $d$ ranges from $0.7$ microns to $10$ nm, the electric power density increases by more than four orders of magnitude, whereas the efficiency improves from $3\%$ to $43\%$. Such improvement is consistent with the improvement in the current and voltage as shown in Fig. \ref{fig:Figure3}.}

\section{Performance of thin-film near-field TPVs}\label{section:Performance}

\par{In the previous section we demonstrated that the near-field photonic interaction between the cell and thermal emitter yields a significant enhancement in luminescence efficiency, and we quantified this enhancement with the parameter $\alpha_\mathrm{NR}$ (Eq. \ref{eq:15}). From Fig. \ref{fig:Figure3}(b), the blackbody model predicts that at small vacuum gaps the considered near-field TPV system operates close to the radiative limit ($Q_\mathrm{e}=1$). The narrowband model as well as our numerical results, however, show that the radiative limit is in fact \textit{never} reached, even for vacuum gaps as small as $d=10$ nm. In this section, we explain this notion and highlight that the near-field enhancement in luminescence efficiency does \textit{not} suggest a suppression of non-radiative recombination. In fact, to the contrary, we show that non-radiative recombination becomes very large when a TPV is operated in the near-field.}

\par{In the previous sections we assumed that non-radiative recombination is approximated by $R(V)=R_\mathrm{o} e^{qV/kT_\mathrm{C}}$ (Eqs. \ref{eq:1}-\ref{eq:3}) \cite{DetailedBalance1961}. Such contributions are included in the numerical calculations shown with black circles, red solid and dashed curves in Figs. \ref{fig:Figure3}, \ref{fig:Figure4}, corresponding, respectively, to fluctuational electrodynamics, the narrowband and the blackbody model. Nevertheless, the non-radiative current of realistic PV cells deviates from the $\sim e^{qV/k T_\mathrm{C}}$ voltage dependence.}

\par{The most prominent non-radiative recombination mechanisms in low-band gap materials are Auger recombination and SRH recombination. Since SRH is strongly dependent on the quality of the crystal and can be suppressed by improving material quality, we only account for Auger recombination here. The Auger non-radiative current is expressed by:
\begin{equation}\label{eq:17}
R(V)=(C_\mathrm{p}p(V)+C_\mathrm{n}n(V))(n(V)p(V)-n_\mathrm{i}^2)t_\mathrm{c},
\end{equation}
where $C_\mathrm{n}$ and $C_\mathrm{p}$ are the Auger coefficients, $n_\mathrm{i}$ is the intrinsic carrier concentration in the cell, and $n(V)$ and $p(V)$ are the electron and hole densities, respectively, given by:
\begin{equation}\label{eq:18}
n(V)=N_\mathrm{c}e^{-(E_\mathrm{c}-E_\mathrm{F,n})/kT_\mathrm{C}},
\end{equation}
and 
\begin{equation}\label{eq:19}
p(V)=N_\mathrm{v}e^{-(E_\mathrm{F,p}-E_\mathrm{v})/kT_\mathrm{C}}.
\end{equation}
$E_\mathrm{F,n/p}$ are the quasi-Fermi levels for electrons/holes, connected to the applied bias $V$ via $V=E_\mathrm{F,n}-E_\mathrm{F,p}$. $E_\mathrm{F,n}$ is determined via the charge neutrality condition across the semiconductor's thickness. $E_\mathrm{c/v}$ are the conduction/valence band edges, respectively.}

\begin{figure}[]
\centering
\includegraphics[width=0.9\linewidth]{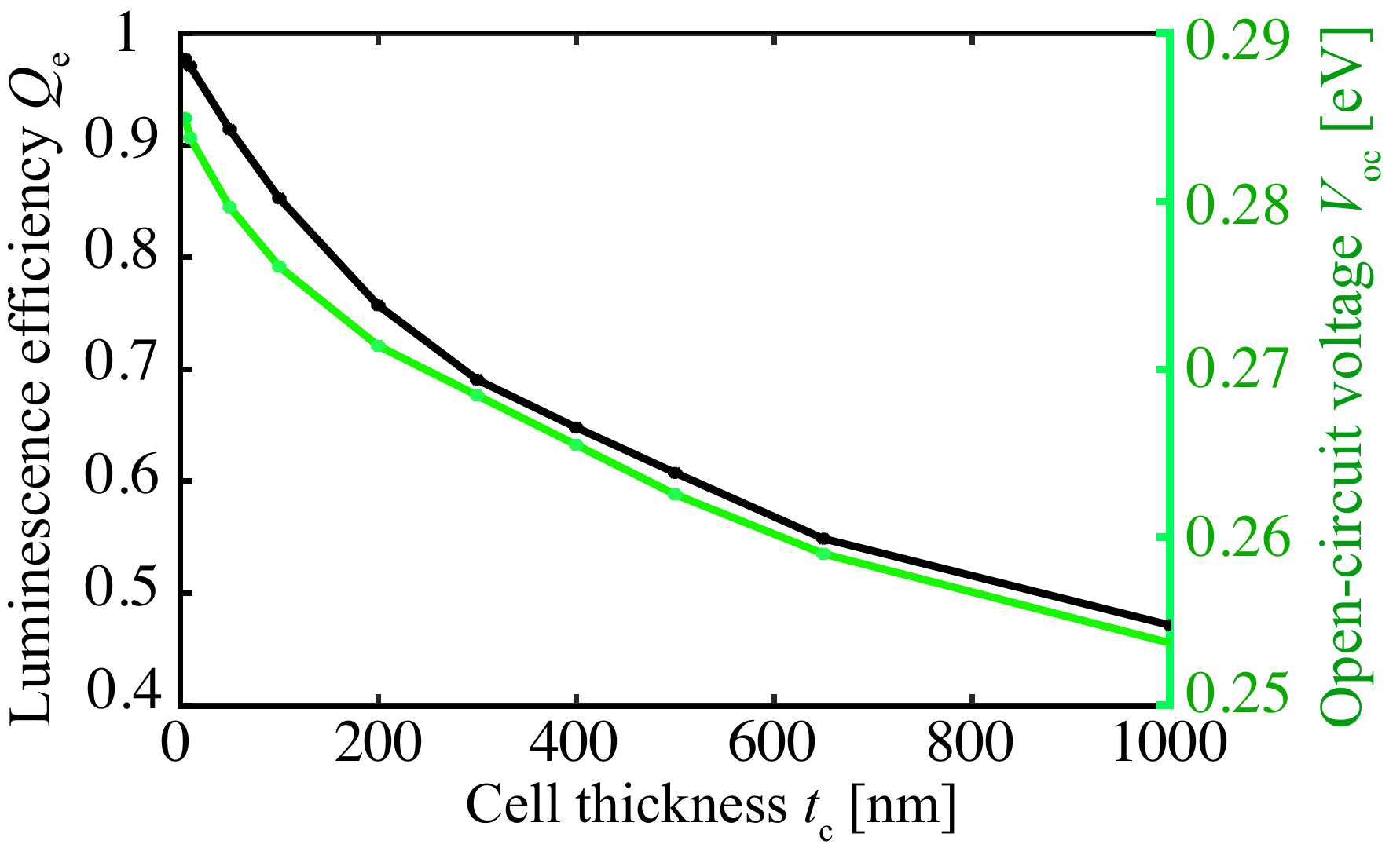}
   \caption{Luminescence efficiency, $Q_\mathrm{e}$, (left axis) and open-circuit voltage, $V_\mathrm{oc}$, (right axis) for a realistic near-field TPV system composed of an InAs PV cell with varying PV cell thickness, $t_\mathrm{c}$, and a $30$ nm ITO plasmonic emitter with plasma frequency $\omega_\mathrm{p}=0.4$ eV, such that it matches the band gap of InAs, at a vacuum gap of $d=10$ nm. The luminescence efficiency is computed at the maximum electrical power point.}
 \label{fig:Figure5} 
\end{figure}

\par{We continue to consider the InAs/ITO system discussed in the previous sections (Fig. \ref{fig:Figure2}(a)), with the difference that, here, the realistic non-radiative recombination model of Eq. \ref{eq:17} is considered. For InAs, we have $C_\mathrm{n}=C_\mathrm{p}=2.26 \times 10^{-27}$ $\mathrm{cm}^{6}$/s \cite{Nonradiative_InAs}. The intrinsic carrier density of InAs is given by $n_\mathrm{i}=\sqrt{N_\mathrm{c}N_\mathrm{v}}e^{\hbar\omega_\mathrm{g,InAs}/(2kT_\mathrm{C})}$, where the effective density of states in the conduction band and valence band, respectively, are $N_\mathrm{c}=8.7\times10^{16}$ $\mathrm{cm}^{-3}$ and $N_\mathrm{v}=6.6\times10^{18}$ $\mathrm{cm}^{-3}$ \cite{Levinshtein_Semiconductors, LIOFEE_1998}. The spectra $\xi_\mathrm{C/H}(\omega,\beta)$ of Eq. \ref{eq:5} are computed rigorously via fluctuational electrodynamics \cite{CHEN2018163}, similar to the results shown with the black circles in Figs. \ref{fig:Figure3}, \ref{fig:Figure4}.}

\par{The results of the calculations with the realistic Auger recombination (Eq. \ref{eq:17}) are shown with the black crosses in Figs. \ref{fig:Figure3}(b), (c), Fig. \ref{fig:Figure4}. Here, the luminescence efficiency is computed as $Q_\mathrm{e}=J_\mathrm{o}(V)/(R(V)+J_\mathrm{o}(V))$ where $J_\mathrm{o}(V)=J_\mathrm{o}e^{qV/kT_\mathrm{C}}$. With Auger recombination, $Q_\mathrm{e}$ is voltage-dependent since $R(V)$ and $J_\mathrm{o}(V)$ have different voltage dependency. Here, we plot the luminescence efficiency at the voltage for which the maximum extracted electrical power density occurs. One can observe a qualitative agreement between the realistic results that consider Auger recombination and the narrowband model that considers $R(V)=R_\mathrm{o,iNaS}e^{qV/kT_\mathrm{C}}$. For the rest of this section, the results are computed with Auger recombination.}

\par{From Eq. \ref{eq:17} it can be seen that Auger recombination is linearly dependent on the cell thickness, hence a thin cell will experience reduced non-radiative losses. To understand the selection of a PV cell as thin as $t_\mathrm{c}=20$ nm, we first analyze the TPV performance as the thickness of the InAs cell varies, while the vacuum gap is fixed. Here, we set $d$ at $10$ nm.  In Fig. \ref{fig:Figure5} we display the dependence of the luminescence efficiency, $Q_\mathrm{e}$ (left $y$-axis), and open-circuit voltage, $V_\mathrm{oc}$ (right $y$-axis), on the thickness of the cell, $t_\mathrm{c}$. The decrease in both $Q_\mathrm{e}$ and $V_\mathrm{oc}$ as the thickness of the cell increases originates from the volumetric nature of Auger recombination (Eq. \ref{eq:17}). This significantly increases non-radiative losses described by $|V_\mathrm{oc,nrad}|$ (Eq. \ref{eq:3}), as $t_\mathrm{c}$ increases. In contrast, the radiative properties that affect $V_\mathrm{oc,rad}$ do not change drastically as the thickness of the cell is varied.}

\par{From Fig. \ref{fig:Figure5}, a thin PV cell exhibits improved luminescence efficiency and $V_\mathrm{oc}$. As discussed in the introduction, the heat-to-electricity conversion efficiency is thermodynamically connected to the open-circuit voltage. Hence, to maximize efficiency, we set the thickness of the InAs PV cell at $t_\mathrm{c}=20$ nm, which is within current fabrication capabilities. We optimize the thickness of the ITO thermal emitter such that efficiency is maximized, for which we obtain an emitter thickness of $10$ nm. The results that follow pertain to this optimized thin-film near-field TPV system.}

\par
\begin{figure}[]
\centering
\includegraphics[width=0.9\linewidth]{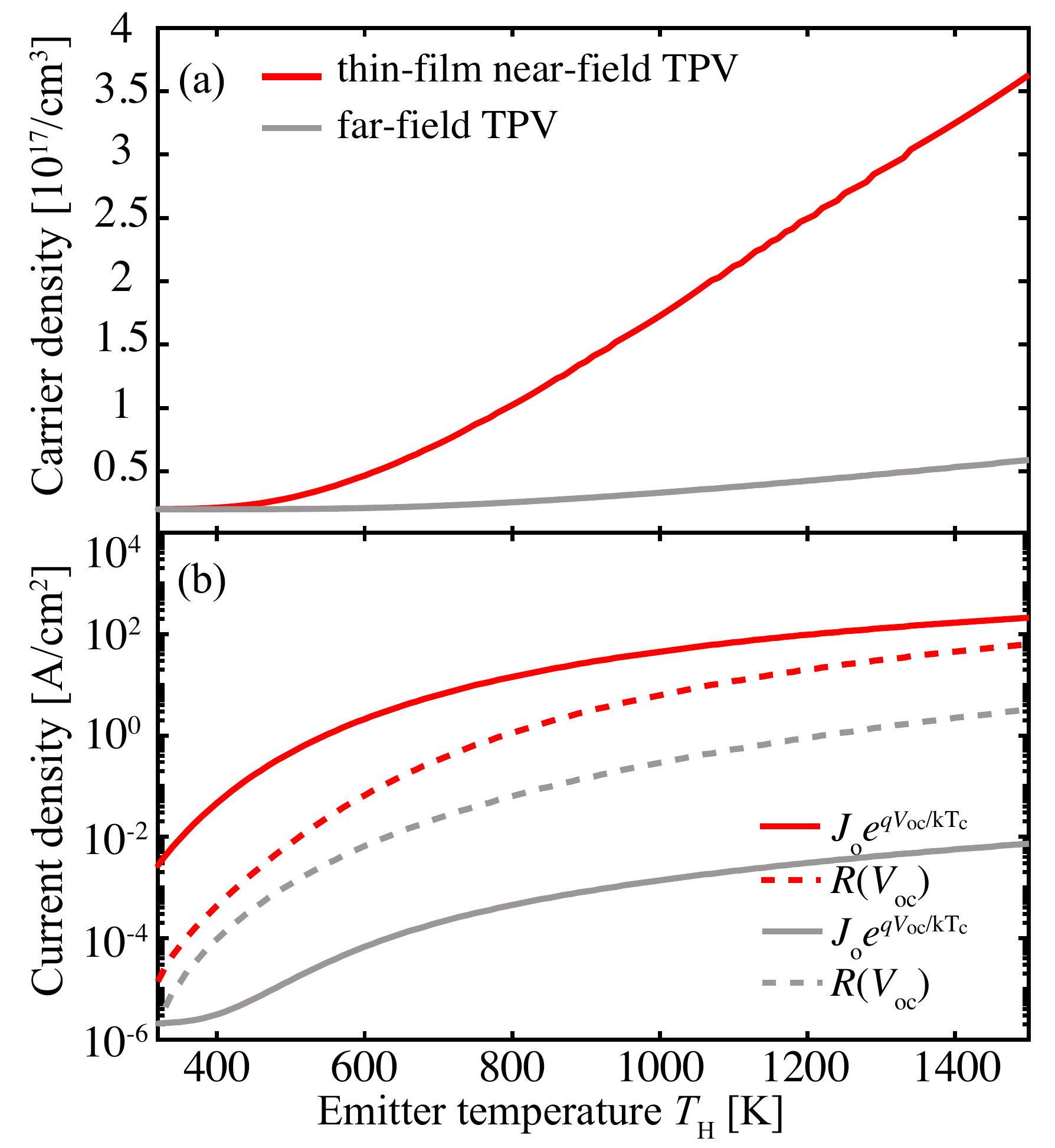}
   \caption{(a) Carrier density and (b) luminescence current and non-radiative current for a thin-film near-field TPV system composed of a $10$ nm ITO emitter and a $20$ nm InAs PV cell (red curves) for a vacuum gap of $d=10$ nm, optimized to maximize efficiency, as compared to a thick ($400$ nm) InAs / ($30$ nm) ITO far-field TPV system for a vacuum gap of $d=10$ $\mu$m. The carrier density and current density are computed at $V=V_\mathrm{oc}$. The far-field TPV system is \textit{not} optimized. Solid curves in panel (b) pertain to the luminescence current, $J_\mathrm{o}$, whereas dashed curves correspond to the Auger current density, $R(V_\mathrm{oc})$, as given by Eq. \ref{eq:17}.}
 \label{fig:Figure6} 
\end{figure}

\begin{figure}[]
\centering
\includegraphics[width=0.9\linewidth]{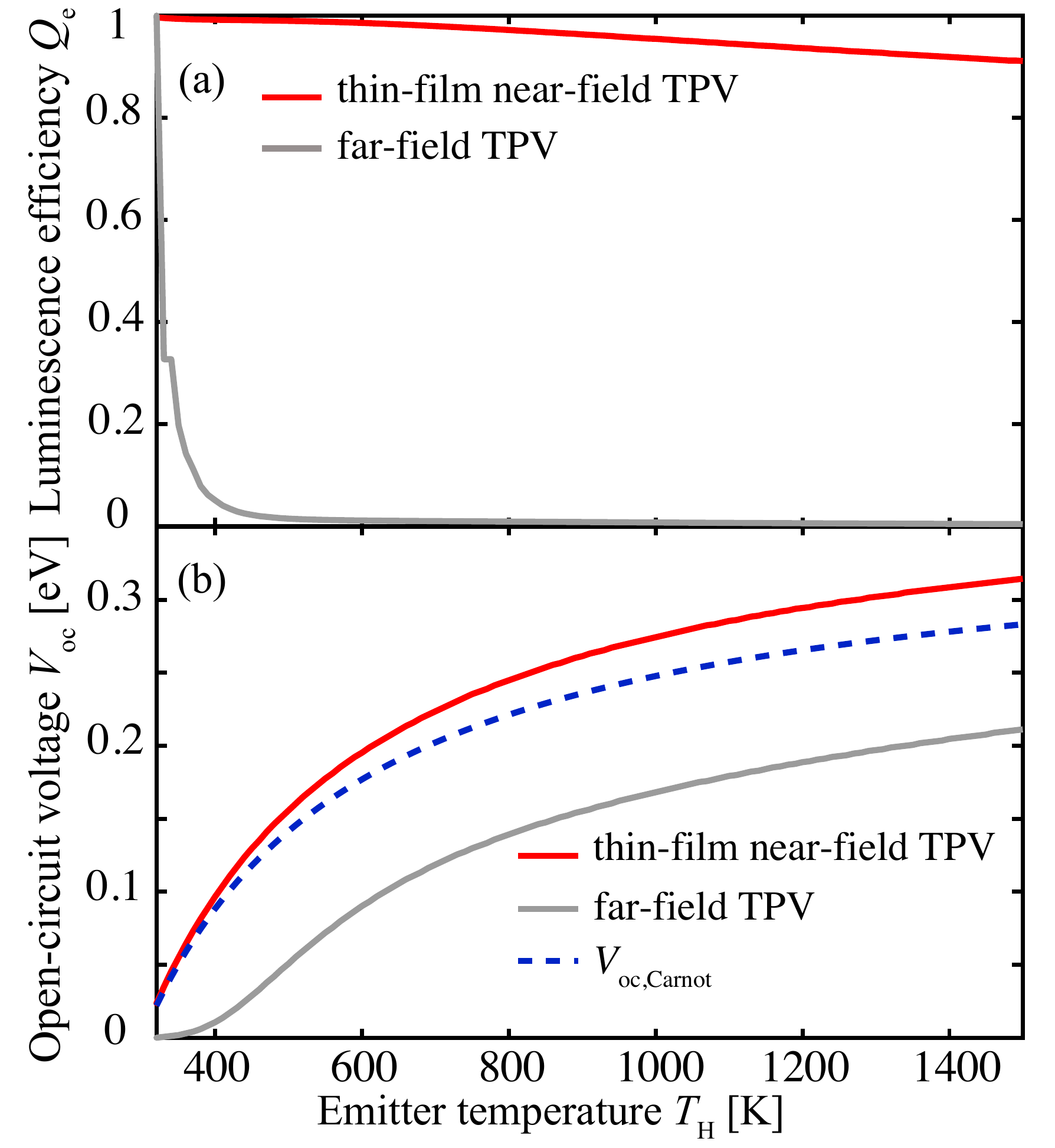}
   \caption{(a) Luminescence efficiency, $Q_\mathrm{e}$, and open-circuit voltage, $V_\mathrm{oc}$, as a function of the emitter temperature, $T_\mathrm{H}$ for the thin-film ($20$ nm) InAs near-field TPV cell and its far-field counterpart, as described in Fig. \ref{fig:Figure6}. The luminescence efficiency is computed at the maximum electrical power point. The dashed blue curve in panel (b) is the product $V_\mathrm{oc,Carnot}=\hbar\omega_\mathrm{g}(1-T_\mathrm{C}/T_\mathrm{H})$.}
 \label{fig:Figure7} 
\end{figure}

\begin{figure}[]
\centering
\includegraphics[width=0.9\linewidth]{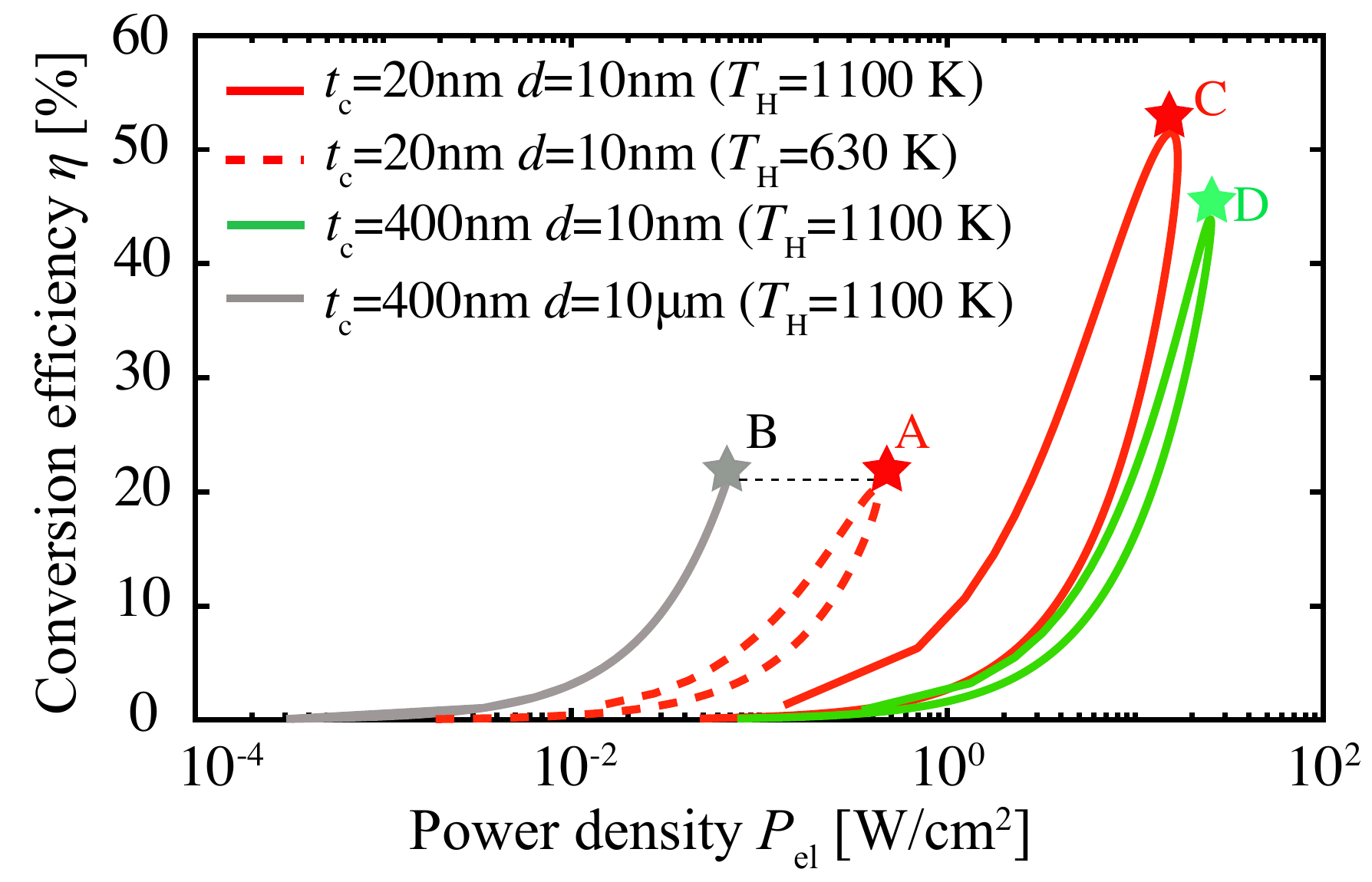}
  \caption{Interplay between extracted electrical power density and conversion efficiency for (A) a near-field $20$ nm-thick InAs TPV cell at a vacuum gap of $d=10$ nm, at an emitter temperature of $T_\mathrm{H}=630$ K, shown with the dashed red curve, (B) a far-field $400$ nm InAs TPV cell at a vacuum gap of $d=10$ $\mu$m, at an emitter temperature of $T_\mathrm{H}=1100$ K, shown with the grey curve, (C) the same thin-film near-field TPV as in (A) for $T_\mathrm{H}=1100$ K, with the red solid curve, and (D) a near-field $400$ nm-thick InAs TPV cell at a vacuum gap of $d=10$ nm at an emitter temperature of $T_\mathrm{H}=1100$ K optimized to maximize extracted electrical power density \cite{Zhao_NFTPV}, shown with the green curve.}
 \label{fig:Figure8} 
\end{figure}

{As a benchmark, we compare the response of this near-field TPV system to its far-field counterpart, for which the vacuum gap is set to $d=10$ $\mu$m, the ITO emitter is $30$ nm thick, and the thickness of the cell is $400$ nm to maximize photon absorption as in \cite{Zhao_NFTPV}. We note that, although not optimized, this far-field TPV system roughly approximates the blackbody model in the far-field (Eq. \ref{eq:6}), since the far-field thermal emission of ITO resembles that of a blackbody, and so does that of InAs for frequencies above its band gap. This case is shown with the grey curves in Figs. \ref{fig:Figure6}-\ref{fig:Figure8}.}

\par{To better understand the interplay between the voltage-dependent luminescence current ($J_\mathrm{o}e^{qV/k T_\mathrm{C}}$), and the Auger current, $R(V)$ (Eq. \ref{eq:17}), we first show in Fig. \ref{fig:Figure6}(a) the carrier density in the InAs cell, while sweeping the emitter temperature, $T_\mathrm{H}$.  In both near-field (red curve) and far-field (grey curve) cases, as $T_\mathrm{H}$ increases, the carrier density increases significantly, as a consequence of the exponential increase in the pump current, $J_\mathrm{e}$ (see Eq. \ref{eq:4}). By comparing the two curves, however, one can see that the carrier density is significantly greater in the near-field as compared to the far-field, even though the near-field TPV cell is $20$ times thinner than its far-field counterpart. This suggests that, in fact, non-radiative Auger recombination is also significantly increased in the near-field (Eqs. \ref{eq:17}-\ref{eq:19}).}

\par{This is shown in Fig. \ref{fig:Figure6}(b) for the near-field thin-film TPV (red dashed curves) and the far-field one (grey dashed curve), where the Auger current is evaluated at $V=V_\mathrm{oc}$. At this voltage, the Auger current is at its maximum. As expected, the Auger recombination rate in the near-field is larger than that in the far-field TPV system by one to two orders of magnitude. However, the luminescence current of the near-field TPV system (red solid curve), evaluated here as $J_\mathrm{o}e^{qV_\mathrm{oc}/kT_\mathrm{C}}$, is significantly larger than that of the far-field TPV (grey solid curve) by four orders of magnitude. This has been explained analytically in Section \ref{section:Vnrad}, where we showed that the luminescence current increases significantly in the near-field. These results provide further insights into the strong external luminescent enhancement. In comparisons with the far-field system, operating in the near-field in fact enhances non-radiative recombination, however, it provides a far larger enhancement of radiative recombination, hence an enhancement of the external luminescent efficiency.}

\par{In Fig. \ref{fig:Figure7}(a), we plot the external luminescence efficiency evaluated at the maximum power point, for the near-field thin-film TPV (red curve) and its far-field counterpart (grey curve), as a function of the emitter temperature. The near-field TPV system exhibits a near-unity luminescence efficiency that significantly surpasses the luminescence of the far-field TPV system. The latter remains below 0.1 for most emitter temperatures. In the far-field system, the external luminescent efficiency has a much stronger temperature dependence since it is dominated by the contributions from the Auger recombination which is also temperature-dependent. In the limit $T_\mathrm{H}\rightarrow 300$ K, the luminescence efficiency in both near-field and far-field TPV systems approaches unity, since Auger non-radiative losses are negligible when $T_\mathrm{H}$ approaches $T_\mathrm{C}$ ($300$ K).}

\par{In Fig. \ref{fig:Figure7}(b) we show the open circuit voltage, $V_\mathrm{oc}$, for the thin-film near-field TPV (red curve) and for its far-field counterpart (grey curve), as a function of the emitter temperature. As expected from the difference between these two systems in terms of luminescence efficiency (Fig. \ref{fig:Figure7}(a)), the $V_\mathrm{oc}$ of the near-field TPV significantly exceeds that of the far-field TPV. It is noteworthy that the $V_\mathrm{oc}$ of the near-field TPV exceeds the term $V_\mathrm{oc,Carnot}=\hbar\omega_\mathrm{g}(1-T_\mathrm{C}/T_\mathrm{H})$, shown here with the blue dashed curve. This effect is attributed to the remarkable advantages of operating in the near-field, namely suppression or radiation leakage (Section \ref{section:Vrad}) and enhancement of luminescence efficiency (Section \ref{section:Vnrad}). As we discussed in the previous sections, surpassing $V_\mathrm{oc,Carnot}$ in solar PV systems is extremely challenging due to imperfect light trapping (Eq. \ref{eq:10}) as well as poor luminescence extraction, hence $V_\mathrm{oc,Carnot}$ is often viewed as an upper bound \cite{PolmanAtwater_highefficiency}. Even in near-field TPV systems, previous literature has also considered $V_\mathrm{oc,Carnot}$ as an upper bound \cite{Ilic_TPV2012}. We note, however, that $V_\mathrm{oc,Carnot}$ does \textit{not} actually represent a thermodynamic constraint on the $V_\mathrm{oc}$ \cite{Hirst_Vocmax2011,Uwe_Voc_2014}. With proper design as we show here, $V_\mathrm{oc,Carnot}$ can be surpassed even in the presence of realistic Auger recombinations. Finally, we note that, as expected from Eq. \ref{eq:12}, the difference between the $V_\mathrm{oc}$ of the near-field TPV and $V_\mathrm{oc,Carnot}$ in Fig. \ref{fig:Figure7}(b) is slightly smaller than $kT_\mathrm{C}\mathrm{ln}(T_\mathrm{H}/T_\mathrm{C})$. The results discussed in Fig. \ref{fig:Figure7}, namely the enhancement in luminescence efficiency and consequent increase in $V_\mathrm{oc}$ in the near-field, explain the good performance previously reported, in terms of both efficiency and power density, for near-field TPV systems \cite{Ilic_Graphene2012, Zhao_NFTPV, Papadakis_2019TPV}.}

\par{In Fig. \ref{fig:Figure8}, we discuss the interplay between extracted electrical power density and the conversion efficiency, as the voltage is tuned from $0$ to $V_\mathrm{oc}$. The voltage at which maximum efficiency is achieved corresponds to the asterisks denoted with letters A, B, C, D. As discussed above, this point differs from $V=V_\mathrm{oc}$ in practical cells. The red dashed curve pertains to the $20$ nm thin-film InAs near-field TPV system as discussed above, for $T_\mathrm{H}=630$ K, whereas the grey curve corresponds to its far-field equivalent, also discussed above, at an emitter temperature of $T_\mathrm{H}=1100$ K. The equal height of points A and B shows that the near-field TPV system yields the same conversion efficiency as its far-field counterpart at nearly half of the required emitter temperature. At the same time, the extracted electrical power density is significantly improved in the near-field. This increase in power density in the near-field was widely considered as a main motivation for operating TPV systems in the near-field \cite{Greffet_NFTPV2006, Zhao_NFTPV}. Yet, Fig. \ref{fig:Figure8} shows that one can additionally significantly decrease the emitter temperature with respect to far-field TPV systems while maintaining good conversion efficiency. Having a lower emitter temperature is practically advantageous. For example, it allevates the difficulty of maintaining material stabiliby at high teperatures. Also, for low-grade waste heat recovery, operating at a lower emitter temperature is essential.}

\par{With the red solid curve and point C in Fig. \ref{fig:Figure8} we show that the same near-field TPV, when operating at the same temperature as the far-field one, namely $T_\mathrm{H}=1100$ K, additionally provides a significant improvement in conversion efficiency with respect to the far-field. The conversion efficiency corresponding to point C is $51.5\%$, which refers to a $143\%$ improvement with respect the far-field TPV (point B). This result is attributed to the significant enhancement of $V_\mathrm{oc}$ in the near-field (Fig. \ref{fig:Figure7}), which is thermodynamically connected to conversion efficiency. This efficiency improvement occurs in addition to an expected increase in power density. The power density is proportional to the current of the cell, which, as can be seen from Eqs. \ref{eq:7}, \ref{eq:9}, scales as $1/d^2$, where $d$ is the TPV vacuum gap.}

\par{Finally, we compare between optically thin near-field TPV cells and optically thick ones. We therefore examine the case of the same material system, but for an InAs cell of thickness $t_\mathrm{c}=400$ nm, as initially proposed in \cite{Zhao_NFTPV}, while the vacuum gap remains at $d=10$ nm. We show the performance of this system with the green curve and point D in Fig. \ref{fig:Figure8}, for an emitter temperature of $1100$ K, as in point C that refers to $t_\mathrm{c}=20$ nm. As can be seen, an ultra-thin cell ($t_\mathrm{c}=20$ nm) at the same $T_\mathrm{H}$ provides a $20\%$ improvement in conversion efficiency with respect to a thick cell ($400$ nm). The reduction in the extracted electrical power density in the thin-film cell (point C-$P_\mathrm{el,C}=16.2$ W/$\mathrm{cm}^2$) with respect to the optically thicker cell (point D-$P_\mathrm{el,D}=25$ W/$\mathrm{cm}^2$) is not significant, considering that point C refers to a cell that is $20$ times thinner than the one of point D. We note additionally that, in practical TPV systems, an extracted electrical power density beyond roughly $10$ W/$cm^{2}$ does not necessarily improve performance because large electrical power density increases electrical losses due to non-radiative recombination, as well as losses in the external wires that connect a practical TPV system to a load, in addition to increasing the temperature of the cell. Last, we note that the power density and conversion efficiency of the TPV systems representing both points C and D in Fig. \ref{fig:Figure8} significantly outperform currently available thermoelectric generators \cite{Thermoelectric_Effifiency, Thermoelectric_PowerDensity,Thermoelectric_PowerDensity2}.}

\section{Conclusion}\label{section:Conclusion}

\par{In conclusion, we presented the theory of thin-film near-field TPV systems that are advantageous in terms of the open-circuit voltage and conversion efficiency as compared to their far-field counterparts. Our theory accurately predicts the performance metrics of near-field TPV systems in terms of both current and voltage, and consequently power density and conversion efficiency. This is to be contrasted with previous analyses that were mainly carried out with numerical simulations based on fluctuational electrodynamics.}

\par{We highlighted two important aspects of TPV operation in the near-field. First, we showed that, due to an increase in the photon density of states near the cell, one obtains an enhancement in the radiative recombination rate. We therefore demonstrated analytically (Eqs. \ref{eq:14}, \ref{eq:15}) that, despite the significant increase in non-radiative recombination in the near-field as compared to the far-field, the external luminescence efficiency is considerably improved. This effect is practically important in particular for low-temperature TPV energy conversion, where low-band gap semiconductors are usually required but typically suffer from large non-radiative losses such as Auger recombination \cite{Kaifeng_Cooling_2015,Francoeur_NFTPV2017}. Furthermore, since non-radiative recombination is volumetric, we considered an ultra-thin cell for minimizing non-radiative losses in terms of $V_\mathrm{oc}$. We therefore showed that very large $V_\mathrm{oc}$ and efficiencies can be obtained with small penalty in the  extracted electrical power density as compared to previously proposed designs \cite{Greffet_NFTPV2006,Ilic_TPV2012,Zhao_NFTPV,Papadakis_2019TPV,Francoeur_NFTPV2011,Francoeur_NFTPV2017,Ideal_NFTPV_Jabob}. Such thin-film near-field TPV systems are therefore promising candidates for efficient TPV energy conversion at significantly lower temperatures as compared to previous proposals with far-field TPV systems.}

\par{As a numerical example, we demonstrated that a near-field TPV composed of a $20$ nm InAs cell and an ITO plasmonic emitter yields an open-circuit voltage that can exceed $\hbar\omega_\mathrm{g}(1-T_\mathrm{C}/T_\mathrm{H})$, which has been often considered as an upper bound on the open-circuit voltage of various PV and TPV systems. Additional factors that will play a role in the performance of a realistic TPV system, beyond the ones considered here, include the increase in the cell temperature as the pump current increases, non-idealities of the back-side reflector, and the effects of electrical contacts on radiative exchange between the emitter and the cell.}

\begin{acknowledgments}
The authors declare no competing financial interest.  We acknowledge the support from the Department of Energy ``Photonics at Thermodynamic Limits'' Energy Frontier Research Center under Grant No. DE-SC0019140. G.T.P. acknowledges the TomKat Postdoctoral Fellowship in Sustainable Energy at Stanford University and Prof. Souzanna Papadopoulou for fruitful discussions.\end{acknowledgments}

\section{\label{SUP}Supplemental Information}
\par{Here, we provide details about the calculation of the luminescence current in Eq. \ref{eq:9} of the main text. This current pertains to the ``narrowband'' near-field emission model that we have introduced. Starting from Eqs. \ref{eq:4}, \ref{eq:5} and \ref{eq:8}, the luminescence current is given by:
\begin{equation}\label{eq:9sup1}
J_\mathrm{o,nf,NR}=\frac{q}{4\pi^2}\frac{\rho^2 \Gamma^2}{2 d^2} \mathrm{I}_\mathrm{NF} e^{-\hbar\omega_\mathrm{g}/kT_\mathrm{C}}.
\end{equation}
The term $\mathrm{I}_\mathrm{NF}$ is given by
\begin{equation}\label{eq:9sup2}
\mathrm{I}_\mathrm{NF}=\int_{0}^{\infty} \frac{e^{-\mu\omega}}{\Gamma^2+\omega^2} d\omega,
\end{equation}
where $\mu=\hbar/(kT_\mathrm{C})$. This integral is given by \cite{gradshteyn2007}:
\begin{equation}\label{eq:9sup3}
\mathrm{I}_\mathrm{NF}=\frac{1}{\Gamma}[\mathrm{ci}(\mu\Gamma)\mathrm{sin}(\mu\Gamma)-\mathrm{si}(\mu\Gamma)\mathrm{cos}(\mu\Gamma)],
\end{equation}
where $\mathrm{ci}(x)=-\int_{x}^{\infty} \frac{\mathrm{cos}t}{t}dt$ and $\mathrm{si}(x)=-\int_{x}^{\infty} \frac{\mathrm{sin}t}{t}dt$. In the limit $\hbar\Gamma\ll kT_\mathrm{C}$ these functions can be approximated by:
\begin{equation}\label{eq:9sup4}
\mathrm{ci}(x)=\gamma+\mathrm{ln}(x)+\frac{-0.25x^2+\mathrm{O}(x^6)}{1+1.16 \times10^{-2} x^2}\approx \gamma+\mathrm{ln}(x)
\end{equation}
and
\begin{equation}\label{eq:9sup5}
\mathrm{si}(x)=-\frac{\pi}{2}+\frac{x+\mathrm{O}(x^3)}{1+10^{-2} x^2}\approx -\frac{\pi}{2}+x
\end{equation}
where $\gamma \approx 0.57722$ is the Euler-Mascheroni constant \cite{Integral_arxiv}. Hence, we obtain $\mathrm{I}_\mathrm{NF}=\pi/(2\Gamma)$, therefore the luminescence current becomes that of Eq. \ref{eq:9} of the main text.}


\begin{thebibliography}{10}

\bibitem{Global_Energy}
``Global energy transformation: A roadmab to 2050,'' {\em International
  Renewable Energy Agency}, 2018.

\bibitem{DetailedBalance1961}
W.~Shockley and H.~J. Queisser, ``Detailed balance limit of efficiency of pn
  junction solar cells,'' {\em Journal of Applied Physics}, vol.~32, no.~3,
  pp.~510--519, 1961.

\bibitem{Harder_2003}
N.-P. Harder and P.~Wurfel, ``Theoretical limits of thermophotovoltaic solar
  energy conversion,'' {\em Semiconductor Science and Technology}, vol.~18,
  pp.~S151--S157, apr 2003.

\bibitem{Record2020_TPV_Forrest}
D.~Fan, T.~Burger, S.~McSherry, B.~Lee, A.~Lenert, and S.~R. Forrest,
  ``Near-perfect photon utilization in an air-bridge thermophotovoltaic cell,''
  {\em Nature}, vol.~586, pp.~237--241, 2020.

\bibitem{DOE2008}
``Waste heat recovery: Technology and opportunities in {U}.{S}. industry,''
  {\em U.S. Department of Energy, Industrial Technologies Program}, 2008.

\bibitem{Zhao_NFTPV}
B.~Zhao, K.~Chen, S.~Buddhiraju, G.~Bhatt, M.~Lipson, and S.~Fan,
  ``High-performance near-field thermophotovoltaics for waste heat recovery,''
  {\em Nano Energy}, vol.~41, pp.~344 -- 350, 2017.

\bibitem{Yablonovich_PNASTPV2019}
Z.~Omair, G.~Scranton, L.~M. Pazos-Out{\'o}n, T.~P. Xiao, M.~A. Steiner,
  V.~Ganapati, P.~F. Peterson, J.~Holzrichter, H.~Atwater, and E.~Yablonovitch,
  ``Ultraefficient thermophotovoltaic power conversion by band-edge spectral
  filtering,'' {\em Proceedings of the National Academy of Sciences}, vol.~116,
  no.~31, pp.~15356--15361, 2019.

\bibitem{Lipson_TPV_experiment}
G.~R. Bhatt, B.~Zhao, S.~Roberts, I.~Datta, A.~Mohanty, T.~Lin, J.-M. Hartmann,
  R.~St-Gelais, S.~Fan, and L.~Michal, ``Integrated near-field
  thermo-photovoltaics for heat recycling,'' {\em Nature Comm.}, vol.~11,
  p.~2545, 2020.

\bibitem{Reddy_NFTPV2018}
A.~Fiorino, L.~Zhu, D.~Thompson, R.~Mittapally, P.~Reddy, and E.~Meyhofer,
  ``Nanogap near-field thermophotovoltaics,'' {\em Nature Nano}, vol.~3, no.~9,
  pp.~806--811, 2018.

\bibitem{Noda_TPV_2019}
T.~Inoue, T.~Koyama, D.~D. Kang, K.~Ikeda, T.~Asano, and S.~Noda, ``One-chip
  near-field thermophotovoltaic device integrating a thin-film thermal emitter
  and photovoltaic cell,'' {\em Nano Letters}, vol.~19, no.~6, pp.~3948--3952,
  2019.


\bibitem{Yablonovitch_Good_LED2012}
O.~D. {Miller}, E.~{Yablonovitch}, and S.~R. {Kurtz}, ``Strong internal and
  external luminescence as solar cells approach the shockley–queisser
  limit,'' {\em IEEE Journal of Photovoltaics}, vol.~2, no.~3, pp.~303--311,
  2012.

\bibitem{Park_NFTPV2008}
K.~Park, S.~Basu, W.~King, and Z.~Zhang, ``Performance analysis of near-field
  thermophotovoltaic devices considering absorption distribution,'' {\em
  Journal of Quantitative Spectroscopy and Radiative Transfer}, vol.~109,
  no.~2, pp.~305 -- 316, 2008.
\newblock The Fifth International Symposium on Radiative Transfer.

\bibitem{Zhao_thermophotonic2018}
B.~Zhao, P.~Santhanam, K.~Chen, S.~Buddhiraju, and S.~Fan, ``Near-field
  thermophotonic systems for low-grade waste-heat recovery,'' {\em Nano
  Letters}, vol.~18, no.~8, pp.~5224--5230, 2018.

\bibitem{Papadakis_2019TPV}
G.~T. Papadakis, S.~Buddhiraju, Z.~Zhao, B.~Zhao, and S.~Fan, ``Broadening
  near-field emission for performance enhancement in thermophotovoltaics,''
  {\em Nano Letters}, vol.~20, no.~3, pp.~1654--1661, 2020.

\bibitem{Francoeur_NFTPV2011}
M.~{Francoeur}, R.~{Vaillon}, and M.~P. {Mengüç}, ``Thermal impacts on the
  performance of nanoscale-gap thermophotovoltaic power generators,'' {\em IEEE
  Transactions on Energy Conversion}, vol.~26, no.~2, pp.~686--698, 2011.

\bibitem{Francoeur_NFTPV2017}
J.~DeSutter, R.~Vaillon, and M.~Francoeur, ``External luminescence and photon
  recycling in near-field thermophotovoltaics,'' {\em Phys. Rev. Applied},
  vol.~8, p.~014030, 2017.

\bibitem{Greffet_NFTPV2006}
M.~Laroche, R.~Carminati, and J.-J. Greffet, ``Near-field thermophotovoltaic
  energy conversion,'' {\em Journal of Applied Physics}, vol.~100, no.~6,
  p.~063704, 2006.

\bibitem{Ilic_TPV2012}
O.~Ilic, M.~Jablan, J.~D. Joannopoulos, I.~Celanovic, and M.~Solja\v{c}i\'{c},
  ``Overcoming the black body limit in plasmonic and graphene near-field
  thermophotovoltaic systems,'' {\em Opt. Express}, vol.~20, no.~S3,
  pp.~A366--A384, 2012.

\bibitem{Kaifeng_Cooling_2015}
K.~Chen, P.~Santhanam, S.~Sandhu, L.~Zhu, and S.~Fan, ``Heat-flux control and
  solid-state cooling by regulating chemical potential of photons in near-field
  electromagnetic heat transfer,'' {\em Phys. Rev. B}, vol.~91, p.~134301, Apr
  2015.

\bibitem{PolmanAtwater_highefficiency}
A.~Polman and H.~A. Atwater, ``Photonic design principles for
  ultrahigh-efficiency photovoltaics,'' {\em Nature Materials}, vol.~11,
  p.~174, 2012.

\bibitem{Rau_Voc_reciprocity_2007}
U.~Rau, ``Reciprocity relation between photovoltaic quantum efficiency and
  electroluminescent emission of solar cells,'' {\em Phys. Rev. B}, vol.~76,
  p.~085303, Aug 2007.

\bibitem{Green_downconversion}
T.~Trupke, M.~A. Green, and P.~Wurfel, ``Improving solar cell efficiencies by
  down-conversion of high-energy photons,'' {\em Journal of Applied Physics},
  vol.~92, no.~3, pp.~1668--1674, 2002.

\bibitem{Markvart_Solar_Heat_Engine}
T.~Markvart, ``Solar cell as a heat engine: energy–entropy analysis of
  photovoltaic conversion,'' {\em physica status solidi (a)}, vol.~205, no.~12,
  pp.~2752--2756, 2008.

\bibitem{Rau_Photon_Recycling}
T.~Kirchartz, F.~Staub, and U.~Rau, ``Impact of photon recycling on the
  open-circuit voltage of metal halide perovskite solar cells,'' {\em ACS
  Energy Letters}, vol.~1, no.~4, pp.~731--739, 2016.

\bibitem{Pendry_NFHT1999}
J.~B. Pendry, ``Radiative exchange of heat between nanostructures,'' {\em
  Journal of Physics: Condensed Matter}, vol.~11, no.~35, pp.~6621--6633, 1999.

\bibitem{Abdallah_betaSPP}
P.~Ben-Abdallah and K.~Joulain, ``Fundamental limits for noncontact transfers
  between two bodies,'' {\em Phys. Rev. B}, vol.~82, p.~121419, Sep 2010.

\bibitem{Max_K_Biehs2010}
S.-A. Biehs, E.~Rousseau, and J.-J. Greffet, ``Mesoscopic description of
  radiative heat transfer at the nanoscale,'' {\em Phys. Rev. Lett.}, vol.~105,
  p.~234301, Dec 2010.

\bibitem{Max_K_Francoer_2011}
M.~Francoeur, M.~P. Meng\"u\ifmmode~\mbox{\c{c}}\else \c{c}\fi{}, and
  R.~Vaillon, ``Coexistence of multiple regimes for near-field thermal
  radiation between two layers supporting surface phonon polaritons in the
  infrared,'' {\em Phys. Rev. B}, vol.~84, p.~075436, Aug 2011.

\bibitem{PolderNFHT}
D.~Polder and M.~Van~Hove, ``Theory of radiative heat transfer between closely
  spaced bodies,'' {\em Phys. Rev. B}, vol.~4, pp.~3303--3314, 1971.

\bibitem{Ideal_NFTPV_Jabob}
S.~Molesky and Z.~Jacob, ``Ideal near-field thermophotovoltaic cells,'' {\em
  Phys. Rev. B}, vol.~91, p.~205435, May 2015.

\bibitem{Joannopoulos_SqueezingNFHT}
A.~Karalis and J.~D. Joannopoulos, ``Squeezing near-field thermal emission for
  ultra-efficient high-power thermophotovoltaic conversion,'' {\em Scientific
  Reports}, vol.~6, p.~28472, 2016.

\bibitem{CHEN2018163}
K.~Chen, B.~Zhao, and S.~Fan, ``Mesh: A free electromagnetic solver for
  far-field and near-field radiative heat transfer for layered periodic
  structures,'' {\em Computer Physics Communications}, vol.~231, pp.~163 --
  172, 2018.

\bibitem{GangChen_NFHTmeasurement2008}
A.~Narayanaswamy, S.~Shen, and G.~Chen, ``Near-field radiative heat transfer
  between a sphere and a substrate,'' {\em Phys. Rev. B}, vol.~78, p.~115303,
  Sep 2008.

\bibitem{Raschke_NFHTReview}
A.~C. Jones, B.~T. O’Callahan, H.~U. Yang, and M.~B. Raschke, ``The thermal
  near-field: Coherence, spectroscopy, heat-transfer, and optical forces,''
  {\em Progress in Surface Science}, vol.~88, no.~4, pp.~349 -- 392, 2013.

\bibitem{NFTPV_review2008}
E.~Tervo, E.~Bagherisereshki, and Z.~Zhang, ``Near-field radiative
  thermoelectric energy converters: a review,'' {\em Frontiers in Energy},
  vol.~12, no.~1, pp.~5--21, 2018.

\bibitem{Abdallah_ReviewNFTPV}
P.~Ben-Abdallah and S.-A. Biehs, ``Harvesting the electromagnetic energy
  confined close to a hot body,'' {\em Zeitschrift für Naturforschung A},
  vol.~74, 2019.

\bibitem{Papadakis_TunableNFHT}
G.~T. Papadakis, B.~Zhao, S.~Buddhiraju, and S.~Fan, ``Gate-tunable near-field
  heat transfer,'' {\em ACS Photonics}, vol.~6, no.~3, pp.~709--719, 2019.

\bibitem{WINSTON1975}
``Principles of cylindrical concentrators for solar energy,'' {\em Solar
  Energy}, vol.~17, no.~4, pp.~255--258, 1975.

\bibitem{Kosten_Angle_Restriction}
E.~Kosten, J.~Atwater, J.~Parsons, A.~Polman, and H.~A. Atwater, ``Highly
  efficient gaas solar cells by limiting light emission angle,'' {\em Light
  Sci. Appl.}, vol.~2, no.~45, 2013.

\bibitem{Yablonovitch_GaAs1992}
I.~Schnitzer, E.~Yablonovitch, C.~Caneau, and T.~J. Gmitter, ``Ultrahigh
  spontaneous emission quantum efficiency, 99.7\% internally and 72\%
  externally, from algaas/gaas/algaas double heterostructures,'' {\em Applied
  Physics Letters}, vol.~62, no.~2, pp.~131--133, 1993.

\bibitem{Javey_TMDs_2016}
M.~Amani, R.~A. Burke, X.~Ji, P.~Zhao, D.-H. Lien, P.~Taheri, G.~H. Ahn,
  D.~Kirya, J.~W. Ager, E.~Yablonovitch, J.~Kong, M.~Dubey, and A.~Javey,
  ``High luminescence efficiency in mos2 grown by chemical vapor deposition,''
  {\em ACS Nano}, vol.~10, no.~7, pp.~6535--6541, 2016.

\bibitem{Nonradiative_InAs}
S.~Krishnamurthy and M.~Berding, ``Full-band-structure calculation of
  shockley-read-hall recombination rates in indium arsenide,'' {\em Journal of
  Applied Physics}, vol.~90, no.~2, pp.~848--851, 2001.

\bibitem{GiantEnhancementSiO2_Reddy2018}
A.~Fiorino, D.~Thompson, L.~Zhu, B.~Song, P.~Reddy, and E.~Meyhofer, ``Giant
  enhancement in radiative heat transfer in sub-30 nm gaps of plane parallel
  surfaces,'' {\em Nano Letters}, vol.~18, no.~6, pp.~3711--3715, 2018.
\newblock PMID: 29701988.

\bibitem{Chen_FanNFTPV2015}
K.~Chen, P.~Santhanam, and S.~Fan, ``Suppressing sub-bandgap phonon-polariton
  heat transfer in near-field thermophotovoltaic devices for waste heat
  recovery,'' {\em Applied Physics Letters}, vol.~107, no.~9, p.~091106, 2015.

\bibitem{Kfaifeng_Auger}
K.~Chen, T.~P. Xiao, P.~Santhanam, E.~Yablonovitch, and S.~Fan,
  ``High-performance near-field electroluminescent refrigeration device
  consisting of a gaas light emitting diode and a si photovoltaic cell,'' {\em
  Journal of Applied Physics}, vol.~122, no.~14, p.~143104, 2017.

\bibitem{Levinshtein_Semiconductors}
M.~Levinshtein, S.~Rumyantsev, and M.~Shur, {\em Handbook Series on
  Semiconductor Parameters}.
\newblock World Scientific, 1996.

\bibitem{LIOFEE_1998}
{\em The Ioffe Physical-Technical Institute of the Russian Academy of Sciences,
  NSM archive}.
\newblock http://www.ioffe.ru/SVA/NSM/Semicond/, 1998.

\bibitem{Hirst_Vocmax2011}
L.~C. Hirst and N.~J. Ekins-Daukes, ``Fundamental losses in solar cells,'' {\em
  Progress in Photovoltaics: Research and Applications}, vol.~19, no.~3,
  pp.~286--293, 2011.

\bibitem{Uwe_Voc_2014}
U.~Rau, U.~W. Paetzold, and T.~Kirchartz, ``Thermodynamics of light management
  in photovoltaic devices,'' {\em Phys. Rev. B}, vol.~90, p.~035211, Jul 2014.

\bibitem{Ilic_Graphene2012}
O.~Ilic, M.~Jablan, J.~D. Joannopoulos, I.~Celanovic, H.~Buljan, and
  M.~Solja\ifmmode \check{c}\else \v{c}\fi{}i\ifmmode~\acute{c}\else
  \'{c}\fi{}, ``Near-field thermal radiation transfer controlled by plasmons in
  graphene,'' {\em Phys. Rev. B}, vol.~85, p.~155422, Apr 2012.

\bibitem{Thermoelectric_Effifiency}
J.~Chen, K.~Li, C.~Liu, M.~Li, Y.~Lv, L.~Jia, and S.~Jiang, ``Enhanced
  efficiency of thermoelectric generator by optimizing mechanical and
  electrical structures,'' {\em Energies}, vol.~10, no.~9, 2017.

\bibitem{Thermoelectric_PowerDensity}
Y.~Zhang, M.~Cleary, X.~Wang, N.~Kempf, L.~Schoensee, J.~Yang, G.~Joshi, and
  L.~Meda, ``High-temperature and high-power-density nanostructured
  thermoelectric generator for automotive waste heat recovery,'' {\em Energy
  Conversion and Management}, vol.~105, pp.~946--950, 2015.

\bibitem{Thermoelectric_PowerDensity2}
Y.~Yu, W.~Zhu, Y.~Wang, P.~Zhu, K.~Peng, and Y.~Deng, ``Towards high
  integration and power density: Zigzag-type thin-film thermoelectric generator
  assisted by rapid pulse laser patterning technique,'' {\em Applied Energy},
  vol.~275, p.~115404, 2020.

\bibitem{gradshteyn2007}
I.~S. Gradshteyn and I.~M. Ryzhik, {\em Table of integrals, series, and
  products}.
\newblock Elsevier/Academic Press, Amsterdam, seventh~ed., 2007.
\newblock Translated from the Russian, Translation edited and with a preface by
  Alan Jeffrey and Daniel Zwillinger, With one CD-ROM (Windows, Macintosh and
  UNIX).

\bibitem{Integral_arxiv}
B.~Rowe, M.~Jarvis, R.~Mandelbaum, G.~M. Bernstein, J.~Bosch, M.~Simet, J.~E.
  Meyers, T.~Kacprzak, R.~Nakajima, J.~Zuntz, H.~Miyatake, J.~P. Dietrich,
  R.~Armstrong, P.~Melchior, S.~Mandeep, and S.~Gill, ``Galsim: The modular
  galaxy image simulation toolkit,'' {\em arXiv:1407.7676 [astro-ph.IM]}.



\end{thebibliography}
\end{document}